\documentclass[12pt,preprint]{aastex}
\usepackage[below]{placeins}
\usepackage[dvips,dvipsnames,usenames]{color}
 \usepackage[bookmarks=false]{hyperref}

\newcommand{\asca}{{\it ASCA}}

\newcommand{\bepposax}{{\it BeppoSAX}}
\newcommand{\xmm}{{\it XMM-Newton}}

\newcommand{\chandra}{{\it Chandra}}

\newcommand{\hst}{\it HST}
\newcommand{\fuse}{\it FUSE}
\newcommand{\civ}{\ifmmode {\rm C}\,{\sc iv} \else C\,{\sc iv}\fi}
\newcommand{\cv}{\ifmmode {\rm C}\,{\sc v} \else C\,{\sc v}\fi}
\newcommand{\cvvi}{\ifmmode {\rm C}\,{\sc v-vi} \else C\,{\sc v-vi}\fi}
\newcommand{\cvi}{\ifmmode {\rm C}\,{\sc vi} \else C\,{\sc vi}\fi}
\newcommand{\nv}{\ifmmode {\rm N}\,{\sc v} \else N\,{\sc v}\fi}
\newcommand{\nvvi}{\ifmmode {\rm N}\,{\sc v-vi} \else N\,{\sc v-vi}\fi}
\newcommand{\nvi}{\ifmmode {\rm N}\,{\sc vi} \else N\,{\sc vi}\fi}
\newcommand{\nvivii}{\ifmmode {\rm N}\,{\sc vi-vii} \else N\,{\sc vi-vii}\fi}
\newcommand{\nvii}{\ifmmode {\rm N}\,{\sc vii} \else N\,{\sc vii}\fi}
\newcommand{\mgix}{\ifmmode {\rm Mg}\,{\sc ix} \else Mg\,{\sc ix}\fi}
\newcommand{\mgixxi}{\ifmmode {\rm Mg}\,{\sc ix-xi} \else Mg\,{\sc ix-xi}\fi}
\newcommand{\mgixxii}{\ifmmode {\rm Mg}\,{\sc ix-xii} \else Mg\,{\sc ix-xii}\fi}
\newcommand{\mgx}{\ifmmode {\rm Mg}\,{\sc x} \else Mg\,{\sc x}\fi}
\newcommand{\mgxi}{\ifmmode {\rm Mg}\,{\sc xi} \else Mg\,{\sc xi}\fi}
\newcommand{\mgxixii}{\ifmmode {\rm Mg}\,{\sc xi-xii} \else Mg\,{\sc xi-xii}\fi}
\newcommand{\mgxii}{\ifmmode {\rm Mg}\,{\sc xii} \else Mg\,{\sc xii}\fi}
\newcommand{\neix}{\ifmmode {\rm Ne}\,{\sc ix} \else Ne\,{\sc ix}\fi}
\newcommand{\nex}{\ifmmode {\rm Ne}\,{\sc x} \else Ne\,{\sc x}\fi}
\newcommand{\neixx}{\ifmmode {\rm Ne}\,{\sc ix-x} \else Ne\,{\sc ix-x}\fi}
\newcommand{\ov}{\ifmmode {\rm O}\,{\sc v} \else O\,{\sc v}\fi}
\newcommand{\ovi}{\ifmmode {\rm O}\,{\sc vi} \else O\,{\sc vi}\fi}
\newcommand{\ovivii}{\ifmmode {\rm O}\,{\sc vi-vii} \else O\,{\sc vi-vii}\fi}
\newcommand{\oviviii}{\ifmmode {\rm O}\,{\sc vi-viii} \else O\,{\sc vi-viii}\fi}
\newcommand{\ovii}{\ifmmode {\rm O}\,{\sc vii} \else O\,{\sc vii}\fi}
\newcommand{\oviiviii}{\ifmmode {\rm O}\,{\sc vii-viii} \else O\,{\sc vii-viii}\fi}
\newcommand{\oviii}{\ifmmode {\rm O}\,{\sc viii} \else O\,{\sc viii}\fi}
\newcommand{\fevii}{\ifmmode {\rm Fe}\,{\sc vii} \else Fe\,{\sc vii}\fi}
\newcommand{\feviixii}{\ifmmode {\rm Fe}\,{\sc vii-xii} \else Fe\,{\sc vii-xii}\fi}
\newcommand{\feviixxiv}{\ifmmode {\rm Fe}\,{\sc vii-xxiv} \else Fe\,{\sc vii-xxiv}\fi}
\newcommand{\feviii}{\ifmmode {\rm Fe}\,{\sc viii} \else Fe\,{\sc viii}\fi}
\newcommand{\feix}{\ifmmode {\rm Fe}\,{\sc ix} \else Fe\,{\sc ix}\fi}
\newcommand{\fex}{\ifmmode {\rm Fe}\,{\sc x} \else Fe\,{\sc x}\fi}
\newcommand{\fexi}{\ifmmode {\rm Fe}\,{\sc xi} \else Fe\,{\sc xi}\fi}
\newcommand{\fexii}{\ifmmode {\rm Fe}\,{\sc xii} \else Fe\,{\sc xii}\fi}
\newcommand{\fexiii}{\ifmmode {\rm Fe}\,{\sc xiii} \else Fe\,{\sc xiii}\fi}
\newcommand{\fexiv}{\ifmmode {\rm Fe}\,{\sc xiv} \else Fe\,{\sc xiv}\fi}
\newcommand{\fexv}{\ifmmode {\rm Fe}\,{\sc xv} \else Fe\,{\sc xv}\fi}
\newcommand{\fexvi}{\ifmmode {\rm Fe}\,{\sc xvi} \else Fe\,{\sc xvi}\fi}
\newcommand{\fexvii}{\ifmmode {\rm Fe}\,{\sc xvii} \else Fe\,{\sc xvii}\fi}
\newcommand{\fexviixx}{\ifmmode {\rm Fe}\,{\sc xvii-xx} \else Fe\,{\sc xvii-xx}\fi}
\newcommand{\fexviixxii}{\ifmmode {\rm Fe}\,{\sc xvii-xxii} \else Fe\,{\sc xvii-xxii}\fi}
\newcommand{\fexviii}{\ifmmode {\rm Fe}\,{\sc xviii} \else Fe\,{\sc xviii}\fi}
\newcommand{\fexix}{\ifmmode {\rm Fe}\,{\sc xix} \else Fe\,{\sc xix}\fi}
\newcommand{\fexx}{\ifmmode {\rm Fe}\,{\sc xx} \else Fe\,{\sc xx}\fi}
\newcommand{\fexxi}{\ifmmode {\rm Fe}\,{\sc xxi} \else Fe\,{\sc xxi}\fi}
\newcommand{\fexxixxiv}{\ifmmode {\rm Fe}\,{\sc xxi-xxiv} \else Fe\,{\sc xxi-xxiv}\fi}
\newcommand{\fexxii}{\ifmmode {\rm Fe}\,{\sc xxii} \else Fe\,{\sc xxii}\fi}
\newcommand{\fexxiii}{\ifmmode {\rm Fe}\,{\sc xxiii} \else Fe\,{\sc xxiii}\fi}
\newcommand{\fexxiiixxiv}{\ifmmode {\rm Fe}\,{\sc xxiii-xxiv} \else Fe\,{\sc xxiii-xxiv}\fi}
\newcommand{\fexxiv}{\ifmmode {\rm Fe}\,{\sc xxiv} \else Fe\,{\sc xxiv}\fi}
\newcommand{\siviii}{\ifmmode {\rm Si}\,{\sc viii} \else Si\,{\sc viii}\fi}
\newcommand{\siix}{\ifmmode {\rm Si}\,{\sc ix} \else Si\,{\sc ix}\fi}
\newcommand{\sixi}{\ifmmode {\rm Si}\,{\sc xi} \else Si\,{\sc xi}\fi}
\newcommand{\sixii}{\ifmmode {\rm Si}\,{\sc xii} \else Si\,{\sc xii}\fi}
\newcommand{\sixiixiv}{\ifmmode {\rm Si}\,{\sc xii-xiv} \else Si\,{\sc xii-xiv}\fi}
\newcommand{\sixiii}{\ifmmode {\rm Si}\,{\sc xiii} \else Si\,{\sc xiii}\fi}
\newcommand{\sixiiixiv}{\ifmmode {\rm Si}\,{\sc xiii-xiv} \else Si\,{\sc xiii-xiv}\fi}
\newcommand{\sixiv}{\ifmmode {\rm Si}\,{\sc xiv} \else Si\,{\sc xiv}\fi}
\newcommand{\six}{\ifmmode {\rm S}\,{\sc ix} \else S\,{\sc ix}\fi}
\newcommand{\sixxii}{\ifmmode {\rm S}\,{\sc ix-xii} \else S\,{\sc ix-xii}\fi}
\newcommand{\sxi}{\ifmmode {\rm S}\,{\sc xi} \else S\,{\sc xi}\fi}
\newcommand{\sxii}{\ifmmode {\rm S}\,{\sc xii} \else S\,{\sc xii}\fi}
\newcommand{\sxiii}{\ifmmode {\rm S}\,{\sc xii} \else S\,{\sc xii}\fi}
\newcommand{\arxi}{\ifmmode {\rm Ar}\,{\sc xi} \else Ar\,{\sc xi}\fi}
\newcommand{\arxixii}{\ifmmode {\rm Ar}\,{\sc xi-xii} \else Ar\,{\sc xi-xii}\fi}
\newcommand{\arxii}{\ifmmode {\rm Ar}\,{\sc xii} \else Ar\,{\sc xii}\fi}
\newcommand{\sx}{\ifmmode {\rm S}\,{\sc x} \else S\,{\sc x}\fi}
  \def\ltsima{$\; \buildrel  <  \over \sim\;$}    \def\simlt{\lower.5ex\hbox{\ltsima}}    
\def\gtsima{$\;      \buildrel      >      \over      \sim      \;$}
\def\simgt{\lower.5ex\hbox{\gtsima}}      

\def\oviii{{\sc O\ viii}}
\def\ovii{{\sc   O\ vii}}

\def\civ{{\sc Civ\/}}
\def\nv{{\sc Nv\/}}
\def\ovi{{\sc Ovi\/}}

\definecolor{gray11}{rgb}{0.11,0.11,0.11}
\definecolor{pink}{rgb}{1,0.75,0.8}
\definecolor{brown}{rgb}{0.65,0.16,0.16}
\definecolor{hotpink}{rgb}{1,0.41,0.71}
\definecolor{deeppink}{rgb}{1,0.08,0.58}
\definecolor{purple3}{rgb}{0.49,0.15,0.8}
\definecolor{orangered3}{rgb}{0.8,0.22,0}
\definecolor{violetred2}{rgb}{0.93,0.23,0.55}
\definecolor{olivedrab1}{rgb}{0.75,1,0.24}
\definecolor{darkolivegreen2}{rgb}{0.74,0.93,0.41}
\definecolor{orange}{rgb}{1,0.65,0}
\definecolor{darkgreen}{rgb}{0,0.39,0}
\definecolor{darkolivegreen}{rgb}{0.33,0.42,0.18}
\definecolor{moccasin}{rgb}{1,0.89,0.71}
\definecolor{dodgerblue}{rgb}{0.12,0.56,1}
\definecolor{darkslateblue}{rgb}{0.28,0.24,0.55}
\definecolor{midnightblue}{rgb}{0.1,0.1,0.44}
\begin{document} 

\title{ THE TWO-PHASE, TWO-VELOCITY IONIZED ABSORBER IN THE SEYFERT 1 GALAXY NGC 5548}
\author{Mercedes Andrade-Vel\'azquez$^{1}$, Yair Krongold$^{1}$, Martin Elvis$^{2}$, Fabrizio Nicastro$^{2,4,5}$, Nancy Brickhouse$^{2}$, Luc Binette$^{1}$, Smita Mathur$^{3}$ \& Elena Jim\'enez-Bail\'on$^{1}$
}

\altaffiltext{1}{Instituto de Astronom\'ia, Universidad Nacional
Aut\'onoma de M\'exico, Apartado Postal 70-264, 04510 M\'exico DF, M\'exico.}
\altaffiltext{2}{Harvard-Smithsonian Center for Astrophysics, 60 Garden
Street, Cambridge MA 02138, USA.}
\altaffiltext{3}{Ohio State University, 140 West 18th Avenue,
Columbus, OH 43210, USA.}  
\altaffiltext{4}{OAR-INAF, Via Frascati, 33, 00040, Monteporzio Catone, RM (Italy)}
\altaffiltext{5}{IESL, Foundation for Research and Technology, 711 10, Heraklion,  
Crete (Greece)}


\begin{abstract}

We present an analysis of X-ray high quality grating spectra of the Seyfert 1 galaxy NGC~5548
using archival \chandra -HETGS and LETGS observations for a total exposure time of
800 ks. The continuum emission (between 0.2-8 keV) is well represented by a
power law ($\Gamma$ = $1.6$) plus a black-body component (kT = $0.1$ keV). We find that the well
known X-ray warm absorber in this source consists of two different outflow
velocity systems. One absorbing system has a velocity of
-$1110\pm{150}$ km s$^{-1}$ and the other of -$490\pm{150} $ km
s$^{-1}$. Recognizing the presence of these kinematically distinct components allows each system
to be fitted independently, each with two absorption components with different ionization levels. The
high velocity system consists of a component with temperature of $2.7\pm0.6\times10^6$ K, $\log U
=1.23$ and another component with temperature of $5.8\pm1.0\times10^5$ K, $\log U =0.67$. The high-velocity, high-ionization component produces absorption by charge states \fexxixxiv, while the high-velocity, low-ionization component produces absorption by
\neixx, \fexviixx, \oviiviii. The low-velocity system required also two absorbing
components, one with temperature of $5.8\pm0.8\times10^5$ K, $\log U
=0.67$, producing absorption by \neixx, \fexviixx, \oviiviii.
The other with lower temperature ($3.5\pm0.35\times10^4$ K), and lower ionization ($\log U =-0.49$);
producing absorption by \ovivii ~and  the {\feviixii} M-shell UTA.
Once these components are considered, the data do not require any further absorbers. In particular, a model consisting of a continuous radial range of ionization structures (as suggested by a previous analysis) is not required.

The two absorbing components in each velocity system are in pressure equilibrium with each other.
This suggests that each velocity system consists of a multi-phase medium. This is the first
time that different outflow velocity systems have been modelled independently in the
X-ray band for this source. The kinematic components and column
densities found from the X-rays are in agreement with the main kinematic components found in
the UV absorber. 
This supports the idea that the UV and X-ray absorbing gas is part of the same phenomenon. NGC 5548
can now be seen to fit in a pattern established for other warm absorbers: 2 or 3
discrete phases in pressure equilibrium. There are no remaining cases of a well studied warm
absorber in which a model consisting of a multi-phase medium is not viable.

\end{abstract}



\section{Introduction \label{intro}}

In $\sim50\%$ Seyferts 1 and quasars there is an interesting phenomenon, the ionized absorber
or warm absorber (WA) (Reynolds 1997; Crenshaw et al. 2003a; Piconcelli et al. 2005).
This phenomenon was observed in the X-ray band for the first time in the spectrum of the quasar MR2251-178 by Halpern (1984) with the Einstein Observatory.  More recently, high resolution observations by \chandra\ and \xmm\ have revealed the presence of blueshifted absorption
lines (v$_{out}\approx-500$ to $-2300$ km s$^{-1}$  with respect to the AGN host rest frame; Collinge et al. 2001; Kaspi et al. 2002), produced by warm ionized gas (T$\sim10^4-10^6$ K; Krongold et al. 2003). These reveal the nature of the WA as an outflowing wind. The absorbing gas usually shows several components with different states of ionization, as it is demonstrated, for
instance, by the different transitions of Oxygen (\ovii,~\oviii),  or the different transitions of Iron (\feviixii~and \fexviixxii). Transitions by \civ, \nv, ~and \ovi~ are observed in both
the X-ray and UV spectra of these sources (with similar outflow velocities, suggesting a connection between the narrow absorption line systems in the UV and the WA (Crenshaw et al. 1999, Kaastra et al. 2000;  Kaspi et al. 2002; Krongold et al. 2003; Steenbrugge et al. 2005).


The possibility that the absorbing components with different ionization states  are in pressure equilibrium with each other, has been suggested by the analysis of several Seyfert 1 galaxies: NCG~3783
(Krongold et al. 2003, Netzer et al. 2003), NCG~985 (Krongold et al. 2005; 2009), NGC  4051 (Krongold et al. 2007), Mrk 279 (Fields et al. 2007).
These results strongly suggests that the absorbers consist of a discrete multi-phase medium.
Is important point out that these studies were carried out by fitting observations using photoionization models. A detailed model for the WA consisting of a thermal multiphase wind,
with a density-size spectrum has been presented by Chelouche \& Netzer (2005); Chelouche  (2008).
Another representation of the absorber  (which includes calculations of line radiative transfer, but does
not fit directly the data) may be a constant pressure medium with a radially stratified density distribution (R\'o\.za\'nska et al. 2006).

\subsection{The Warm Absorber in NGC~5548}
NGC 5548 (classified as a Seyfert 1 with Broad Emission Lines, Full Width Half Maximum (FWHM$_{H{\beta}}$) = 5610 km s$^{-1}$; Turner et al. 1999) is at a redshift $z=0.017175\pm{0.000023}$ (De Vaucouleurs,  1991; the
determination of z was performed with the detection of an HI emission line with negligible error in the velocity, Heckman et al. 1978), and has a luminosity of L$_{2-6KeV} =  (2.0-2.8)\times10^{43}$ erg s$^{-1}$ (Branduardi-Raymont, 1986; Nicastro et al. 2000; present work).

This galaxy has been studied many times in the optical, UV, and
X-ray bands. The optical variability of the source allows detailed reverberation
mapping studies of the Broad Emission Line Region (BELR; e.g. Peterson, 1993), yielding a
central black hole mass of $6.7\pm2.6 \times10^7$ M$_\odot$ (Peterson 2004).
This is $\sim 33$ times larger than, for instance, the NGC 4051 black hole mass (2$\times10^6$ M$_\odot$, Peterson et al. 2004)\footnote{However, note that
Marconi et al. (2008) suggested that  the black hole mass of
Narrow Line Seyfert~1 galaxies (like NGC 4051) is underestimated by a factor
$\sim$5, when  radiation pressure is taken into account in the
virial theorem.}, for which the WA has been extensively studied and
has been described as an accretion disk wind (Krongold et al. 2007). The BELR radius calculated using the H${\beta}$ emission line  for NGC 5548 is 21.2$\pm2.4$ l-d (Kaspi et al. 2000) and the He II BELR is 16.2 l-d (Bottorff 2002). This is $\sim3$ and $\sim5$ times larger, respectively, than the NGC 4051 BELR radius.

The X-ray flux is variable (Branduardi-Raymont, 1989; Nandra et al.
1991; Chiang et al. 2000; Markowitz et al. 2003).  The long-term X-ray and optical
continuum light curves are highly variable and are correlated (Uttley et al. 2003). It has been suggested
that these variations are due to thermal instabilities in the
inner disk (Treves et al. 1988). The short-term X-ray variability
may be more concentrated toward the center of the accretion
flow (Uttley et al.).

Low resolution X-ray studies (e.g. Nicastro et al. 2000, using \bepposax\ satellite
~observations) showed the presence of an ionized absorber, as the data presented transitions consistent with bound-free absorption by H-like and He-like ions of O. The first high resolution
spectrum of NGC 5548, obtained with the \chandra
~Low Energy Transmission Grating Spectrometer (LETGS), confirmed
the presence of the ionized absorber with many absorption lines from
transitions such as \cvi,~\nvii,~\oviiviii,~and \neixx~ (Kaastra et
al. 2000). In the UV band, spectra obtained with the  Goddard High-Resolution
Spectrograph (GHRS) on the Hubble Space Telescope (HST) presents
multiple velocity components, with intrinsic absorption lines such as
Ly$_\alpha$, \civ$\lambda 1549$, and \nv$\lambda 1242$ (Mathur et
al. 1999, Crenshaw et al. 2003b). Five different kinematic  
blueshift components are observed with velocities corresponding to -166, -336,
-530, -667, and -1041 km s$^{-1}$ (components 5 to 1 in Crenshaw et al. 2003b).
Far Ultraviolet Spectroscopic (FUSE)
spectra also show intrinsic \ovi\ and Ly$_{\beta}$ absorption
(Brotherton et al. 2002).

NGC 5548 is the first Seyfert galaxy for which a possible connection between the X-ray and UV absorbers was suggested (Mathur et al. 1995) given the ionization state of
the medium (though a connection has already been suggested for the
quasar 3C 351 by Mathur et al. 1994).  However, the exact nature of this connection in NGC 5548 has been controversial. Brotherton et al. (2002) disputed this conclusion by the discrepancies in the \ovi\ ionic column densities inferred in the two bands. Arav et al. (2003) then found consistency between these columns, once the effects of velocity-dependent covering fraction in the UV band are considered. However using simultaneous \chandra-LETGS and HST-STIS\footnote{(Hubble Space Telescope-Space Telescope Imaging Spectrograph)}  
observations, Crenshaw et al. (2003)b report that the ratio
$\frac{N(\nv)} {N(\civ)}$,  in the UV component 1, is 2.5, which
suggests a gas with low ionization state, implying low \ovii\ and
\oviii\ column densities. Crenshaw et al., then suggested no
relation between the UV and X-ray absorbers. Using \chandra ~ High Energy Transmission Grating Spectrometer (HETGS) and LETGS data, Steenbrugge et al. (2005) proposed again a link between that X-ray and UV absorbers, given the similar 
kinematics seen in the absorption lines in the two bands.  A new study based on 2002 UV spectra suggested that at least a part of
UV component 3 should produce absorption in X-rays range (Crenshaw et al. 2009).
While it is clear that some connection exists, the exact relation  between the UV and X-ray absorbers remains uncertain.

NGC 5548 is perhaps an exception to the idea that the WA is a multi-phase medium possibly in pressure balance (\S \ref{intro}). A continuous radial range of ionization states was suggested for the structure of the WA in this source (Steenbrugge et al. 2005). It should be noted that NGC 5548 has an unusually low Galactic column density (1.6$\times10^{20}$ cm$^{-2}$, Murphy et al. 1996), so the X-ray spectrum remains less absorbed at soft X-ray wavelengths than in most AGNs. This broader wavelength range exposes extra atomic transitions that can provide additional constraints on WA conditions (for instance NGC 3783, has a very well studied absorber, but has a substantially higher Galactic column density than NGC 5548, 9.5$\times10^{20}$ cm$^{-2}$, Murphy et al. 1996).  Thus, it is worthwhile
to test the multi-phase scenario for the absorber in this object.

For these reasons, an analysis of all the seven \chandra\ datasets seems in order.
We re-analyzed archival \chandra\ LETGS and HETGS high resolution
spectra of NGC 5548. Despite previous results that suggest that
only a continuous radial flow could model the absorber (Steenbrugge et
al. 2005), we show that its physical structure is also consistent with a
multi-phase medium. We contrast our results with previous studies of
this source. In section 2 we describe the
observations and data reduction. In section 3 we describe our
data modeling. In \S\ 4 we discuss our results, and in \S\ 5
their implications.

\section{Data Processing \label{d.r.}}

\begin{figure}

\caption{NGC 5548 light curve for the observations analyzed in this
work. Points 2 and 3 correspond to HETGS observations and points 1, and 4-7 to LETG observations.} \label{fig:lc}
\plotone{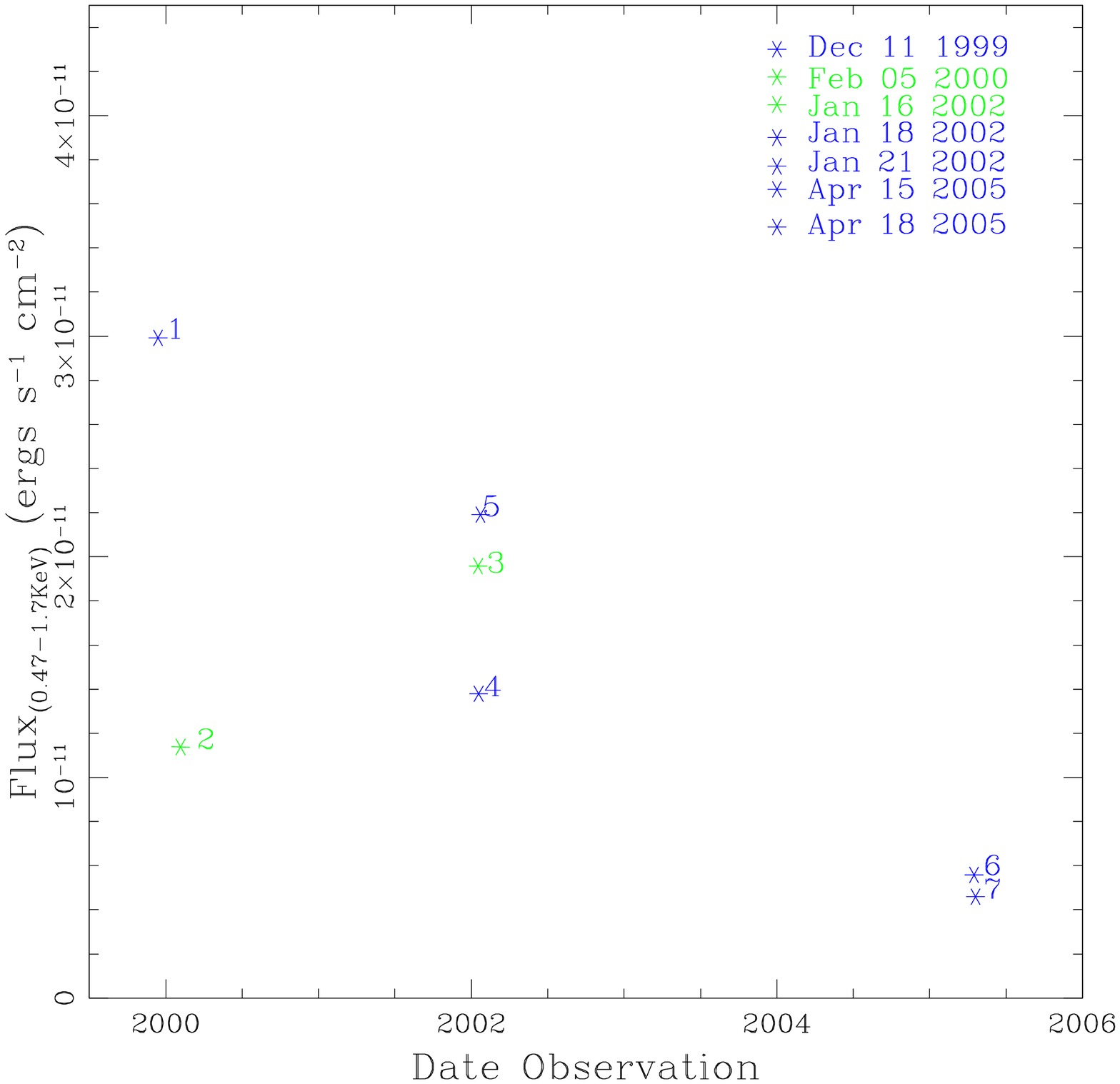}
\end{figure}

NGC 5548 was observed seven times with the \chandra~ X-ray Observatory (Weisskopf et al. 2000) between 1999 and 2005 (see Table \ref{tab:obs}) .
Two observations were carried out with the HETGS (Canizares et al. 2000)
with the light dispersed on to the Advanced CCD Imaging Spectrometer (ACIS; Garmire
et al. 2003). The remaining five observations were
obtained with the LETGS
(Brinkman et al. 2000) and the High Resolution Camara (HRC; Murray et al. 2000). We retrieved
the primary and secondary data products of the observations from the
\chandra\ data archive\footnotemark.\footnotetext{http://cxc.harvard.edu/cda/}
The data reduction was performed with the \chandra\ Interactive Analysis of
Observations software (CIAO v.3.4
\footnote{http://cxc.harvard.edu/ciao3.4/index.html}; Fruscione et al. 2002). We followed the data analysis threads provided by the \chandra\ X-ray Center\footnotemark. Negative and positive first order spectra were extracted and 
response matrices created for each observation. In this paper, we explore the
time-averaged properties of the spectra, focusing on the best possible
determination of the different ionization and velocity components of
the well-known ionized absorber in this AGN. Thus, we co-added the
spectra obtained with each instrument to get the maximum signal to
noise ratio (S/N). This gave a total net exposure time
of 236 ks for the MEG and HEG spectra, and 564 ks for the LEG
spectrum.  The source shows flux differences among observations.  The change
in flux  between the 2 HETGS observations is $\approx$1.7. The change
in flux among the LETGS data is a factor of 6 during the six
years of observations (see  Fig.\ref{fig:lc}).  The change
in flux in short time scales between consecutive (LETGS) observations is
$\approx$1.2-1.6  (see Table \ref{tab:obs}), (for observations separated by 3
days). A time-evolving analysis is postponed to a forthcoming
paper (Andrade-Velazquez et al. in preparation). However, we discuss
the possible effects that these variations in our analysis in Section \ref{effects}.

\footnotetext{http://cxc.harvard.edu/ciao/threads/index.html}

We have analyzed the data in the range between (1.57 to 10.81)~ \AA\ ([1.15
to 7.90] keV) for the HEG,  (2.46 to 24.58)~ \AA\ ([0.50 to 5.04] keV) for MEG, and  (8.85 to 49.15)~ \AA\ ([0.25 to 1.40] keV) for LEG.  All the data were binned by a factor of 2, corresponding to  0.005~\AA, 0.01~ \AA, and 0.025~ \AA\ per bin for the HEG, MEG and LEG, respectively.

\section{Modeling \label{model}}

We have used the CIAO software
Sherpa\footnote{http://cxc.harvard.edu/sherpa} (Freeman et al. 2001)
to perform spectral fits to the data. In all the models described below, the attenuation due to the Galactic absorption was accounted for with a neutral absorber (see \S\ 1). We first focus on the HETGS (HEG and MEG) data. Then, we compare the final HETGS model with the LETGS data, and find a good fit at both short and long wavelengths (see below).

\subsection{ Fitting the HETG data.}
\subsubsection{MODEL A: No Absorption Components.}
Initially, we fit the data with a power law, following previous analysis (Reynolds et al. 1997: Advanced Satellite for Cosmology and Astrophysics -\asca; Nicastro et al. 2000 \bepposax) and a
black-body component (Kaastra $\&$ Barr 1989, Steenbrugge et al. 2005),see Table \ref{tab:models}, Model A. We further added four Gaussians to model the evident emission lines (see Table \ref{tab:EL}\footnote{This Table lists eight emission lines, as a detailed analysis of the spectra \S 3.4  revealed the presence of another four emission lines}).

Three emission lines are resolved, with FWHM between $~$500-700 km s$^{-1}$ (MEG  has an FWHM resolution of $\backsimeq$ 155 ~km s$^{-1}$ at 19~ \AA). These lines are in the
soft band  (consistent with \neix$\lambda$13.69, \ovii$\lambda$ 22.10, and \nvi$\lambda$ 23.27). The short wavelength ($\leqslant 5~ \AA$) region of the spectra shows the presence of the fourth emission line. This corresponds to the commonly found low-ionization ($\leq$ \fexvii) Fe K$_\alpha$
line (6.4 keV, Nandra et al. 2007), and is located at 1.937~\AA ~(without any
shift from the rest frame of the host galaxy).  Yaqoob et al. (2001) report an EW of 133 eV and a width of 4515$^{+3525}_{-2645}~km s^{-1}$ and suggest a possible origin in the outer Broad Line Region (BLR). We report a full width
half maximum (FWHM) of 2644$\pm $2469 ~km s$^{-1}$ consistent with the
value reported by Yaqoob et al. We find a slightly lower EW than this authors,
EW$\sim$ 92.2$\pm43.9$ eV, however, a detailed analysis of this line is beyond the scope of this paper.

\begin{figure}

\caption{Model A plotted over \chandra-MEG data ( [1.6-24.5] \AA) of NGC 5548. The model (red thick line) consists of a powerlaw and a black-body component plus four strong emission lines (blue labels).
Possible transitions due to the ionized absorber are indicated with red labels.}\label{fig:modA}
\plotone{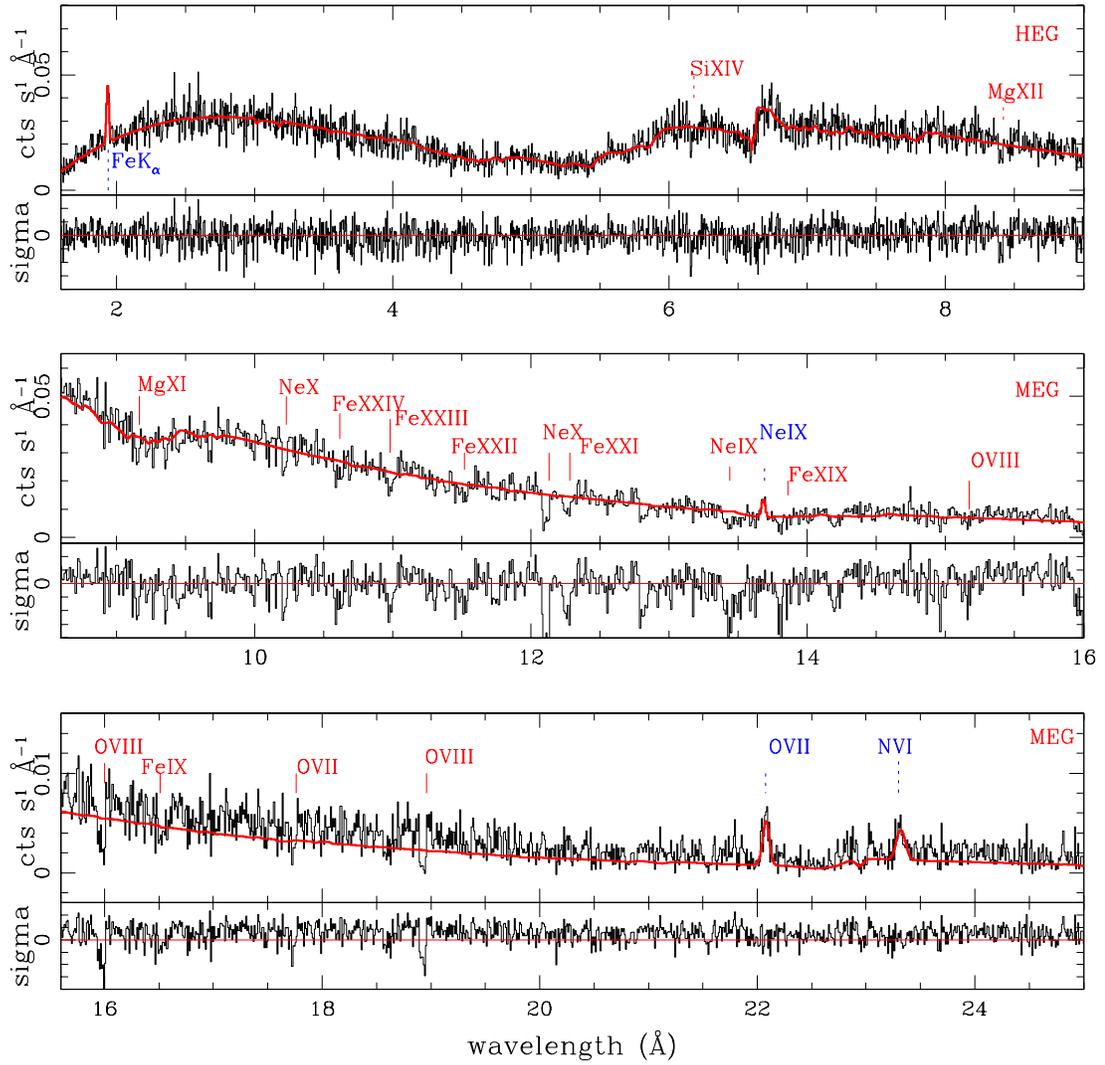}
\end{figure}

Though model A was statistically acceptable ($\chi^2$/d.o.f.= 3813/4112), strong negative residuals between (5-24.5) \AA~are present, consistent with absorption
features by ionized atoms (\fexviixxii, \oviiviii, \neixx, \sixiiixiv, \nvivii,~and \mgxixii)  due to the well-known WA in this source (see Fig.\ref{fig:modA}).

\subsection{ \textbf{Ionized Absorber : fitting with PHASE}}

To model the features produced by the ionized absorber, we used the code PHASE (Krongold et al. 2003). The code has 4 free parameters per absorbing component, namely  1) the ionization parameter at the illuminated face of the absorber $U = \frac{Q(H)}{4\pi r^2 n_H c} $, where  $Q(H)$ is the rate of H ionizing photons, $r$ is the distance to the source, $n_H$ is the
hydrogen number density and $c$ is the speed of light; 2) the hydrogen
equivalent column density $N_{H}$, 3) the outflow velocity of the
absorbing material V$_{out}$ and 4) the micro-turbulent velocity
V$_{turb}$ of the material. Usually, the micro-turbulent velocity is
not left free to vary, because the different transitions due to
ionized gas are heavily blended, or because the different velocity
components are also blended and cannot be resolved (e.g. Krongold et
al. 2003, 2005 for NGC 3783). In the case of NGC 5548, with the HETGS
data resolution, it is possible to distinguish two velocity components
(see below). So, despite the fact that the individual absorption lines
cannot be resolved, we have left the micro-turbulent velocity free to vary (at least
for the strongest absorbing components, see below). PHASE has the advantage of producing a
self-consistent model for each absorbing component, because the code starts without any prior constraint on the column density or population fraction of any ion. We have assumed solar elemental abundances (Grevesse \& Noels 1993). So, given the intrinsic Spectral Energy Distribution (SED, the approximation of this input parameter is
explained in the next subsection) of the source, the column density, and the ionization parameter of the absorbing media, the ionization balance is fixed.

In all the models described below, we have assumed that the continuum source (including the soft excess) is fully covered by the ionized absorbing gas while the emission lines are produced outside the region where the absorbing material is located.

\subsubsection{ \textbf{Spectral Energy Distribution}}

\begin{figure}
\caption{Spectral Energy Distribution used in this analysis (black
solid line) to model  the ionized absorber in the {\em  Chandra}
data of NGC 5548 (the SED is based on our observations and NED data,
see \S \ref{model} for details). Blue labels mark different observational bands (see
\S\ 3). The slopes adopted in each band are marked with red labels
(this slope relates to the photon index as $\Gamma$=s+2).  The
dashed green line represents the SED used by Steenbrugge et al. 2005.}
\label{fig:sed}
\plotone{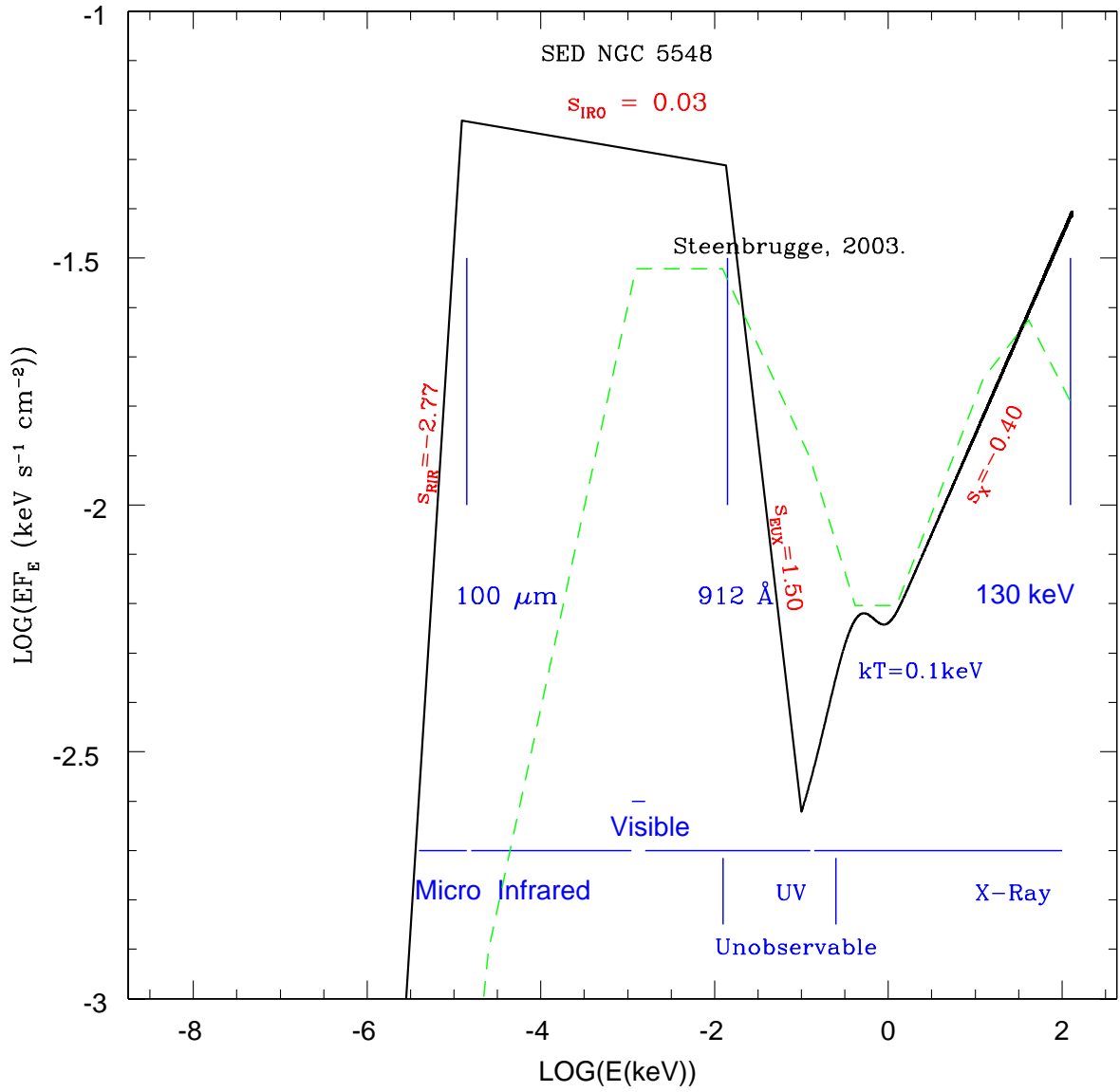}
\end{figure}

The X-ray to radio SED of NGC 5548,  used to calculate
the ionization balance of the gas, was defined in the following way:
Between the radio and the near UV ranges, we used (non simultaneous) data from NED (NASA/IPAC  Extragalactic Database\footnote{http://nedwww.ipac.caltech.edu/}, references
are attached in Appendix \S \ref{app} ). In the range
between 100 $\mu$m and 912 \AA, we performing a linear fit to the
over 100 references of flux measurements found in NED. For the region between (0.25-7.90) keV ( [1.57 - 49.15]~\AA), we used the power law ($\Gamma$) as well as the black-body contribution, as inferred from the fits to the continuum presented in Table \ref{tab:contMH}. For energies
$\geq 8$ keV, we extrapolated the derived $\Gamma$ in our fits up to
130 keV, where a cut-off is observed (Nicastro et al. 2000). Given that the UV SED shape has only a
second order effect on the ionization balance of the X-ray species, as these charge states have
ionization potentials larger than 0.1 keV (see Netzer 1996,
Steenbrugge et al. 2003), we have connected the unobservable region of
the SED between 912 \AA\ and 0.25 keV  with a simple power law (Haro-Corzo et al. 2007). The radio region (at 21 cm) was connected with the infrared (100 $\mu$m) also with a
simple power law.  Figure \ref{fig:sed} shows the final SED used in this work.

\subsubsection{\textbf{Modelling Averaged Spectra}\label{effects}}
An important caveat of averaging the different
datasets is that the the physical properties of the absorbers might
have changed among observations, either because of intrinsec changes or because of changes
induced by the variations in the flux level. We note, that Kaastra et al. 2002 and Steenbrugge et al. 2005
report similar values for the Hydrogen equivalent column density
(within the uncertainties) for the 1999 and 2002 observations (see
Table \ref{tab:obs}).  Then, we do not expect a significative change
in the equivalent Hydrogen column density among the  seven observations. The same is true
for the outflow velocity of the absorbing gas. 

However, the flux level changes between the
observations used in this work. Assuming the gas is in photoionization
equilibrium, this would drive changes by the same factors observed in
the ionization parameter (U) of the gas (producing changes in the gas
opacity; Nicastro et al. 1999).  There is a flux variation by a factor
of 1.7 between the HETG observations, thus, the expected variation
between these two observations will be Delta (logU) = 0.23. For the
LETG observations, there is a large change in flux (by a factor of 6
during 6 years; see Fig. \ref{fig:lc}) between
the first and last two observations. In photoionization equilibrium
this would produce a change Delta(logU) =0.77 in the ionization
parameter. Thus, our determination of this parameter in the
co-added data will represent a weighted average over the different
observations.  We note that for the LETG, the 1999 and 2002
observations have the higher S/N and similar fluxes (different by a
factor $\sim 2.19$, and also similar to the HETG observations).
Then, the best fit value of the absorbing gas ionization parameter on 
the LETG co-added spectrum will be highly dominated by these observations
(the inclusion of the 2005 fainter observations contribute to
increase the S/N  by $\sim10\%$).

\subsection{ \textbf{Ionized Absorber}}

\subsubsection{Model B: One Ionized Absorber Component.}

\begin{figure}

\caption{Model B plotted over \chandra-MEG data ( [1.6-25] \AA). This
model (red thick line) includes the same components of model A plus one
absorbing component. Emission and absorption lines are marked as in
Figure \ref{fig:modA}.}\label{fig:modB}
\plotone{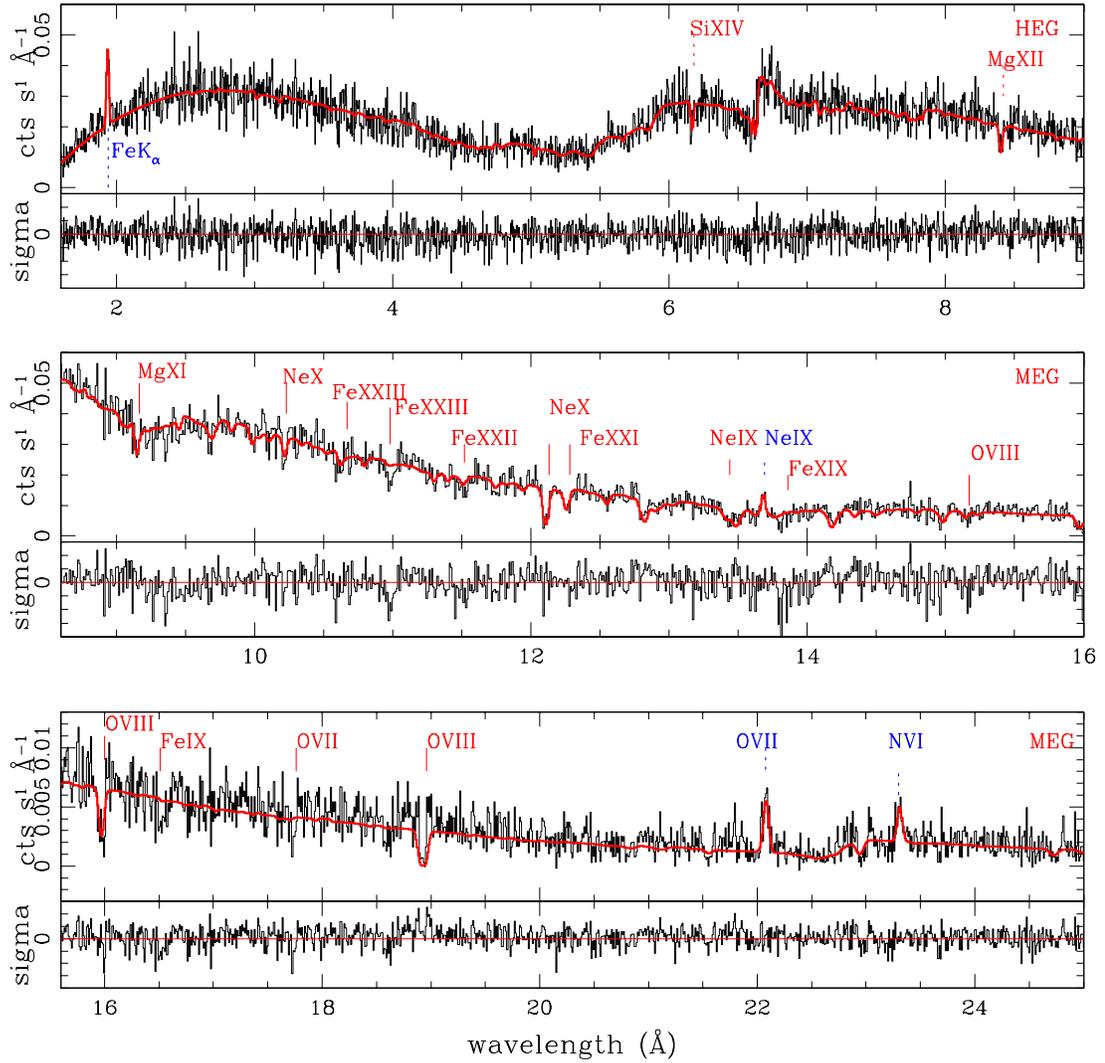}
\end{figure}

\begin{figure}
\caption{Model B and residuals in selected regions of the spectrum showing that the presence of additional absorbing material with higher and lower ionization is required by the data.}\label{fig:modBc}
\plotone{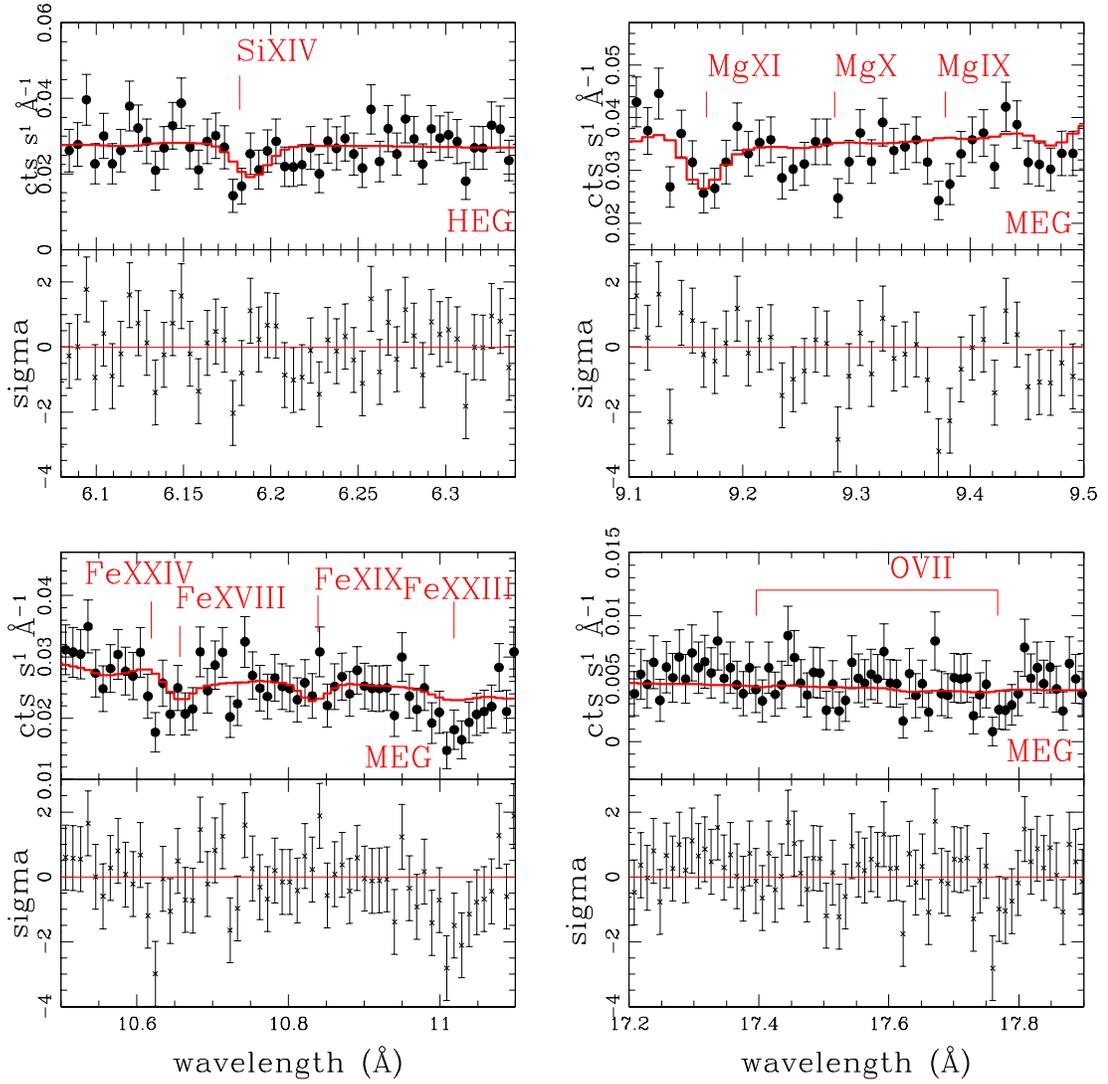}
\end{figure}

We made a new model, B, by including a single ionized absorbing component to
model A. Table \ref{tab:comp} shows the best fit parameters values for this model. The outflow
velocity is $\sim$~ -740$\pm150$ km s$^{-1}$ and could correspond to the UV component 2,
which has a value $\sim$~ -667 km s$^{-1}$. Given the  ionization parameter (logU$_{B}$)
and column density  (logN$_{HB}$) of this component (see Table \ref{tab:models}), this model fits the absorption lines produced by ions such as \fexviixxii, \neixx, \sixiii, \mgxixii,~and \oviii~(see Fig. \ref{fig:modB} ). The fit gives a significant improvement over model A ($\chi^2=3053 $ for 4108 d.o.f.; $\Delta\chi^2 = 760$). An F-test gives a higher than $99.99\%$ confidence level for the presence of the absorber (see Table  \ref{tab:models}). The
absorbing component can reproduce $\sim$ 20 absorption features (including blends) in the
wavelength region (5 - 20)~\AA. However, significant residuals remain at 6.18~\AA, which are probably due to a \sixiv~transition ($\lambda$6.182), and at 10.62~\AA ~and 11.01~\AA, probably corresponding to transitions by \fexxiv($\lambda$10.619) and \fexxiii($\lambda$11.019), see Figure \ref{fig:modBc}; indicating the possible presence of another absorbing  component of gas with higher ionization. Strong residuals found close to 16.5~\AA, could correspond to inner transitions of \feix, further suggesting the possible presence of another absorber with lower ionization state.

\begin{figure}

\caption{Detailed line profile for several absorbing features that are
not well fitted by only one broad absorbing component.  The dotted lines correspond to the outflow velocity of the UV components (Crenshaw et
al. 2003). Significant residuals are found with outflow velocity
consistent with UV component 1.}\label{fig:modBb}
\plotone{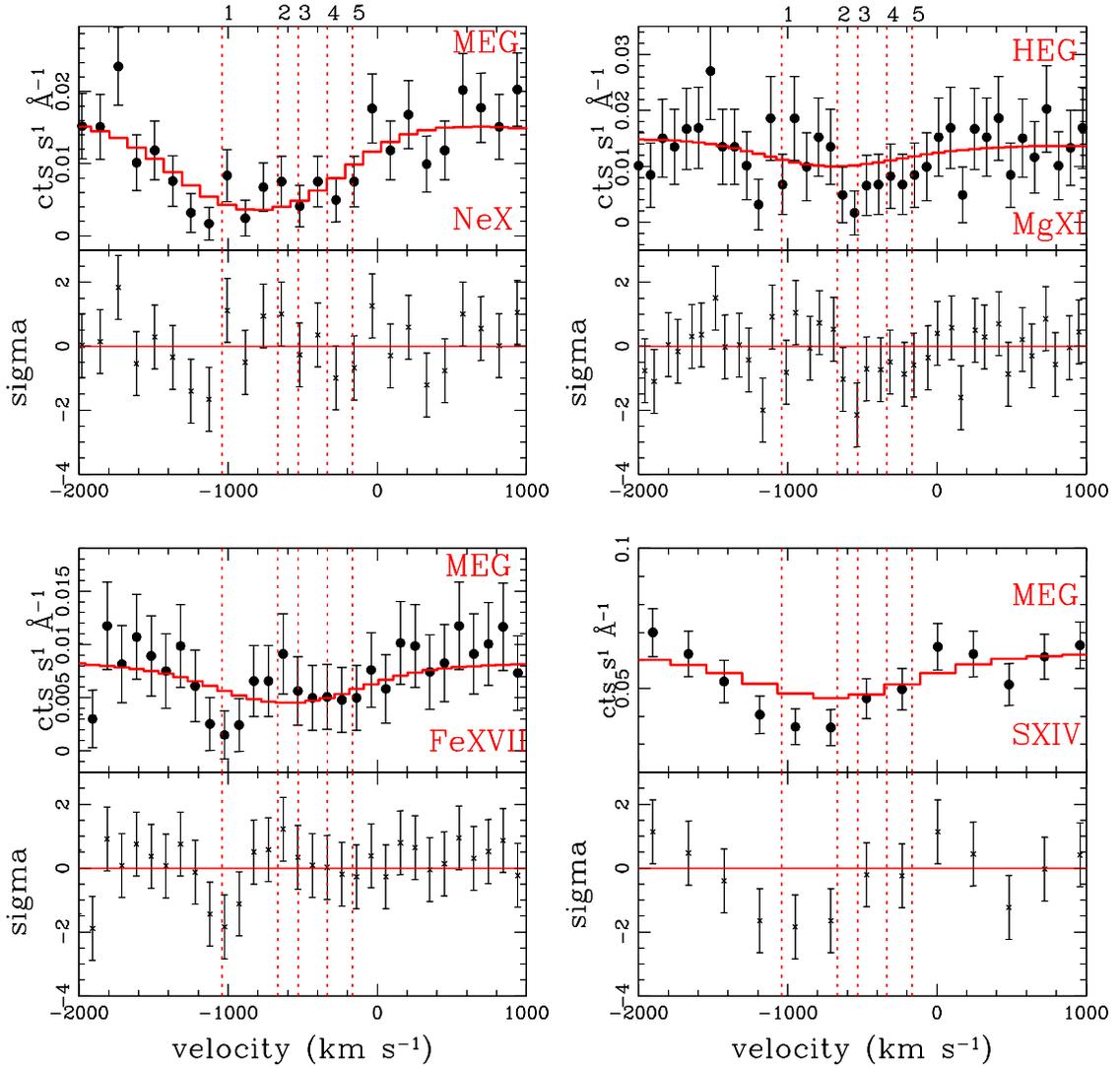}
\end{figure}

In addition, a detailed inspection of the spectrum near the absorption lines fit by this model (\mgxi, \neix,
\fexvii,~and \sixiv) reveals a complex structure in the line profile, with additional absorption with an outflow velocity of $\sim$ -1040 km s$^{-1}$ (see Fig. \ref{fig:modBb}).
These results, implying additional components and complex
velocity structure, are consistent with the analysis of Steenbrugge et
al. (2005). We will model these additional components in the following sections.

\subsubsection{Model C: Two Ionized Absorber Components.}

Model C includes two absorbing components, whose parameters are again
free to vary. This model gives a $\chi^2 =2987$ for 4104 d.o.f., significantly improving
over model B ($\Delta\chi^2 =66$, an F-test, gives a confidence of larger
than $99.99\%$ for the addition of these parameters). The ionization
parameters (logU$_{C1}$ and logU$_{C2}$) are different by almost a factor of 3 (see Table
\ref{tab:comp}).

\begin{figure}

\caption{As Figure \ref{fig:modB}, but for model C, which includes two
absorbing components. This model can fit transitions by both high
and medium ionization level ions such as Fe$_{XXIII-XXIV}$.}\label{fig:modC}
\plotone{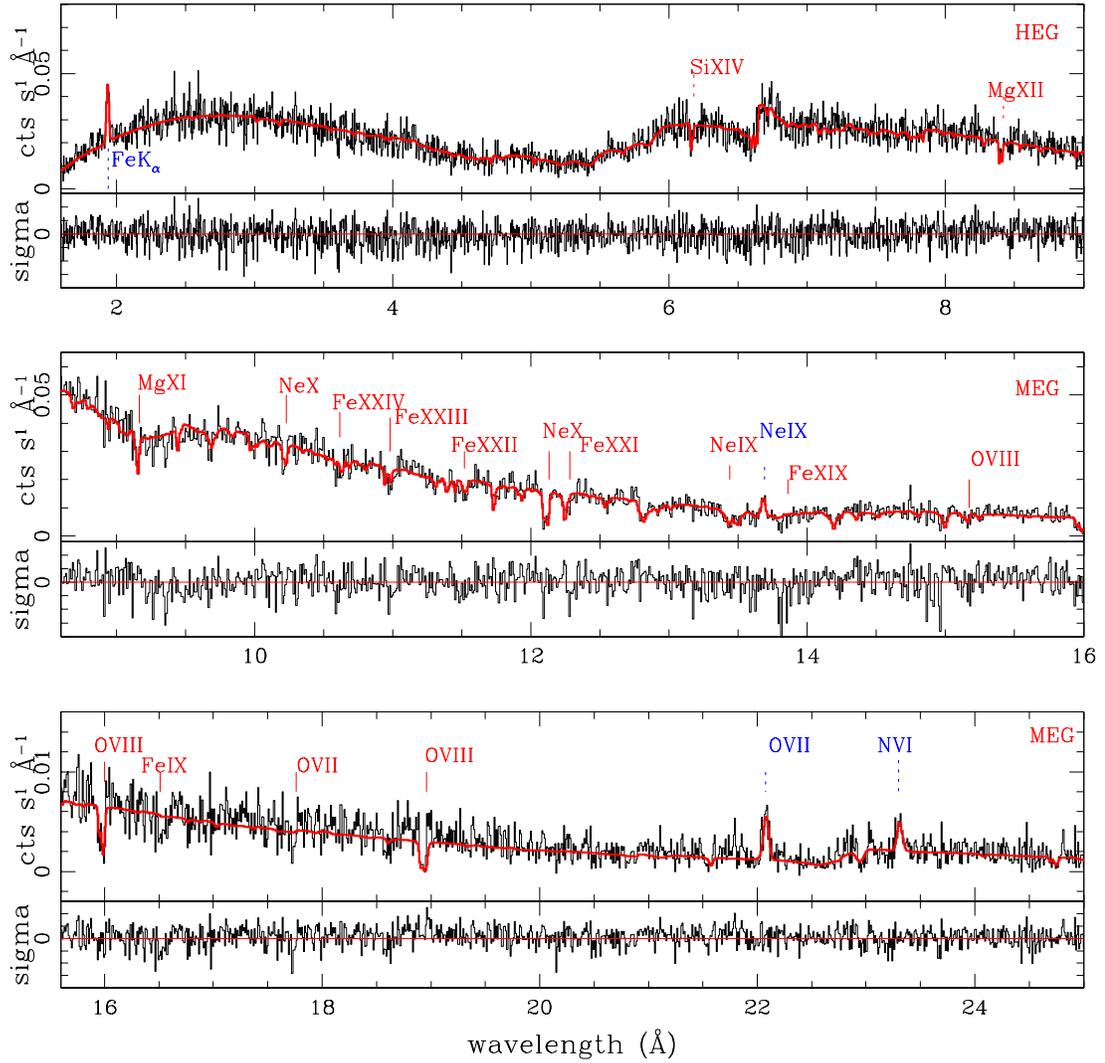}
\end{figure}

The component with lower ionization (with logU$_{C2}$ slightly lower than  logU$_{B1}$) models the transitions already fitted in model B: \sixiiixiv, \neixx, \mgxixii, \fexviixxii,~and \oviii~ (see Fig.
\ref{fig:modC}). The higher ionization component (with logU$_{C1}\sim1.4\times$logU$_{B1}$) contributes weakly to fit these ions, but fits the transitions due to \fexxiiixxiv~(not fitted in
model B). The two components have different outflow velocities (
V$_{outC1}$ =  -1120 km s$^{-1}$ and V$_{outC2}$ = -450 km
s$^{-1}$) . We note that while model B partially fits the \sixiv~
line, the higher ionization component in model C (logU$_{C1}$), which
has also higher outflow velocity, improves the fit to the
\sixiv\ line, i. e. this component fits the
residuals that model B left around 6.18 \AA~ (see Fig.
\ref{fig:modC}). While model C does reproduce, then, the lines
produced by \sixiv~, \fexxixxiv, it cannot model the complex absorption profile observed also for other lines.

\subsubsection{Model D: Three Ionized Absorber Components.}

\begin{figure}
\caption{As Figure \ref{fig:modB}, but for model D (red thick line), which
includes three absorbing components. The contribution of the lower
ionization component is observed in transitions such as Mg$_{IX-XI}$ and O$_{VII}$.}\label{fig:modD}
\plotone{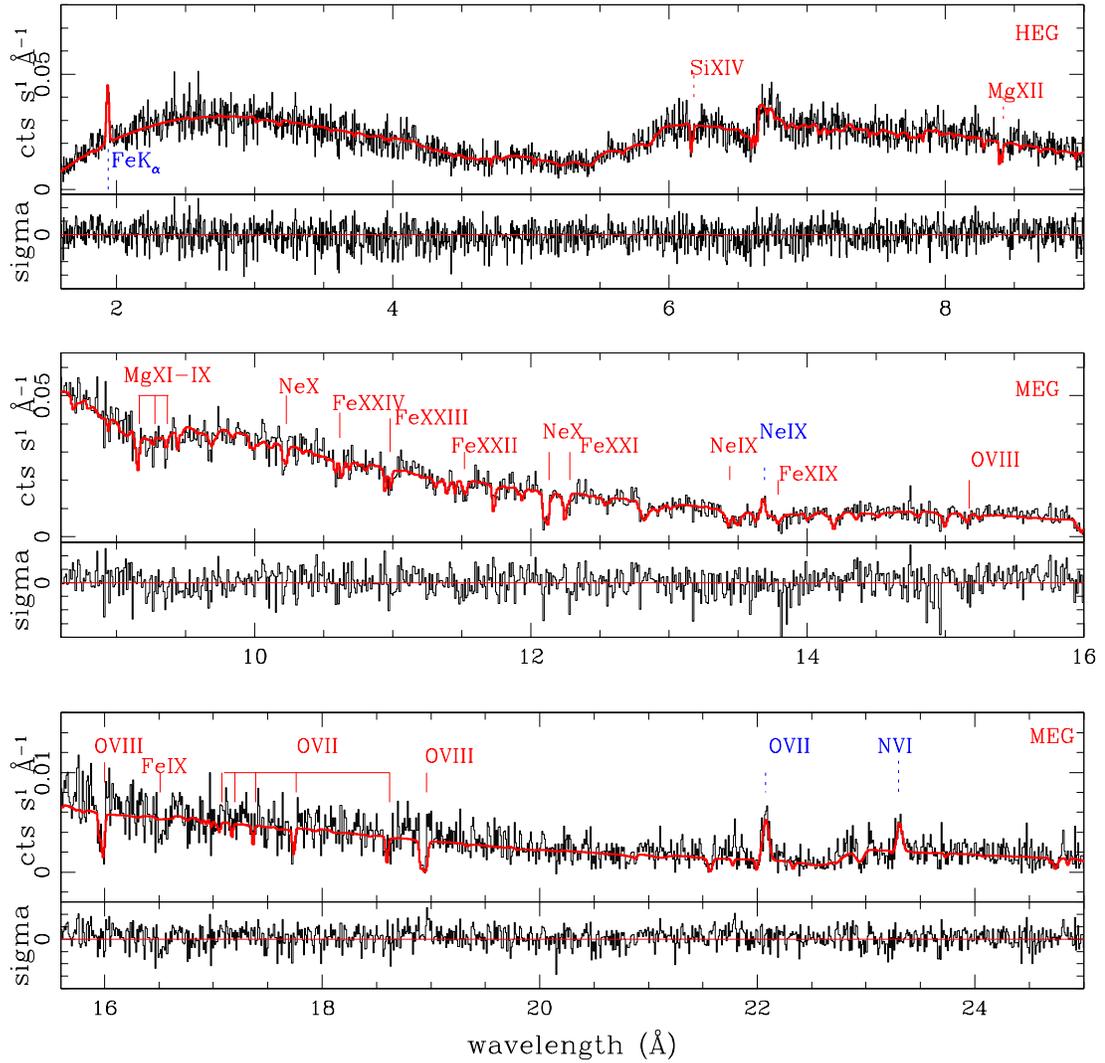}
\end{figure}

Motivated by the line profile residuals, we built a new model, D, by adding a third
absorbing component, whose parameters are again free to vary, as are those of the other two
components. Model D gives a $\chi^2 =2946$ for 4100 d.o.f., improving
significantly model C ($\Delta\chi^2 = 41$, F-test confidence level of $99.99\%$).
All the three absorbing components have different ionization level. Two of the three absorbing components have similar values of logU to those found in model C. The third component  has a much lower ionization level (by at least a factor $\sim14$, see Table \ref{tab:comp}), and  fits the \nvivii, \ovii, \mgixxi~ lines, as well as the \feviixii~ M-shell Unresolved Transitions Array [UTA] (see Fig. \ref{fig:modD}). The three absorbing components group in two different outflow velocity systems. One system,  with outflow velocity in the range between -450$\pm150$ and -590 $\pm150$ km s$^{-1}$, contains the systems with intermediate ( logU$_{D2}$) and low (logU$_{D3}$) ionization level. The other velocity system (V$_{out}$ =  -1116 $\pm$ 150 km s$^{-1}$) is formed by the high ionization component.

We note that model D still leaves some residuals around 16.5~\AA, where an \feix\ transition lies. The
reason for this discrepancy may be due to our use of the abbreviated data to model the UTA (see Krongold et al. 2003 and Behar et. al 2001 for further details).

\subsubsection{Model E: Four Ionized Absorber Components.}

Model D does not solve the problem of the line profile for several
absorption lines with intermediate ionization state, in particular the \fexvii~and \nex~ transitions. Therefore, model E was built by adding a fourth absorbing component to model D.
This case gives a $\chi^2 = 2904$ for 4096 d.o.f. ($\Delta\chi^2 = 42 $), which is statistically better
than that of model D (an F-test shows that the fourth absorber is required at the $99.99\%$ confidence
level). The best-fit parameters value for each absorbing component are
listed in Table \ref{tab:2abHV}, and the global fit to the HETGS data
is shown in Figures \ref{fig:meg} (for MEG data) and \ref{fig:heg}
(for HEG data) .
Model E again consists of two outflow velocity systems (hereafter High
Velocity or HV for the system with V$_{out}$ =  -1110 $\pm$ 150 km
s$^{-1}$, and Low Velocity or LV with V$_{out}$ =  -490 $\pm$ 150 km
s$^{-1}$ ), each having two different ionization
components\footnote{ The velocities for the HV
 and LV systems were calculated taking the average of the two
 components forming each system (see Table \ref{tab:2abHV})}.
The ionization parameter of the high ionization phase in the LV system
(hereafter LV-HIP) is logU=0.67 with a column density of log$N_{H}$= 21.25.  The low ionization phase (LV-LIP) has  log$N_{H}$= 20.74 and logU=-0.49.
The features fitted  by this system are the same ones fitted in model
D by the components similar to that model. The contribution of this
velocity component to the global fit is shown in Figure \ref{fig:LV}.

The HV system has two absorbing components. One of the absorbing
components of this system has a high ionization level (logU=0.671,
hereafter called HV-HIP), and a column density of logN$_{H}$=21.03,
and the other absorber has an even higher ionization level (hereafter
called HV super-high ionization phase or HV-SHIP) with logU=1.23 and
logN$_{H}$=21.73. Figure \ref{fig:HV} shows the contribution of the HV
system to the global fit of the spectrum.
The HV-SHIP fits the same features model D does.

The HV-HIP of this system fits the same absorption
lines as the LV-HIP (given that the ionization parameters of these two
components are quite similar). The difference between these two components lies mainly in
their outflow velocities, finally improving the fit to the line profile (as seen in
Fig. \ref{fig:lines2c}). We note that this result is obtained without
any constraint in the main parameters of the absorbers, which further confirms the presence of the
two velocity systems. Separating the two components was possible given
the excellent resolution of the HETGS, $\approx$155 km s$^{-1}$ (19 \AA) for MEG\footnote{The resolution here and anywhere else is given with
respect to the factor of 2 binning used in this analysis.}.

\subsection{Emission Features in the Spectra of NGC 5548 \label{el}}

In the previous models we fitted four strong emission lines (the Fe
K$_{\alpha}$ line, and the \neix, \ovii,~and \nvi~ forbidden
transitions). However, in the MEG data, there are
four additional emission lines that can be detected with a
significance level $> 2 \sigma$ once the absorbers are properly
modelled. These lines  correspond to the resonant
and intercombination transitions of \ovii, the resonant transition
of \nvi~and \sixiii\ line. We produced a new model
including these lines (model F). The best fit parameters for the eight
absorption lines included to fit the HETG data are given in Table
\ref{tab:EL}. Figure \ref{fig:meg} already includes these features.

Our analysis does not require the inclusion of broad emission
features (as was reported by Steenbrugge et al. 2005).
The forbidden (z), resonant (w), and intercombination (x+y) lines
of He-like ions (Gabriel \& Jordan 1969) can be used to derive the
physical conditions of photoionized plasma (Porquet $\&$ Dubau
2000). In particular, the ratio between the resonant and
intercombination lines serve as a diagnostic of the electronic density
n$_{e}$ (R(n$_{e}$) = $\frac{z}{x+y}$). Using the  R value of the \ovii\ lines
(R(n$_{e}$)= 2.23$\pm1.6$), and following Porquet $\&$ Dubau (their
Figure 8), we find that the gas must be photoionized and must have a
density n$_{e}=2.0\pm0.8\times10^{10} cm^{-3}$. Kaastra et al. (2000)
report an upper limit for n$_{e} <$ 7.0$\times10^{10} cm^{-3}$, with a
R(n$_{e}$) = 2.22$\pm3.45$, which is in agreement with our result.

\subsection{  \textbf{ Extrapolation to LETGS}}

The final model (F) for the ionized absorber in the MEG and HEG
spectra of NGC 5548 can be further tested by extrapolating to the
LETGS data (the LETGS resolution is a factor $\sim$ 2.5 worse than the
MEG resolution at 19 \AA). In doing this, we implicitly assume no
significant opacity variations between the co-added HETGS and LETGS
spectra that may arise from flux variability (Nicastro et al. 1999, Krongold et al. 2005b, 2007). Given that the average flux level in these two spectra differs by only 30\% (Steenbrugge et al. 2005), this is a reasonable assumption \footnote{This is not the case between individual observations, where opacity changes might be present. See Andrade-Velazquez et al. (in preparation) for a full analysis of variability of the absorber to continuum changes.}.

To fit the LEG spectrum, we used model F, with the 4 absorbing
components' parameters fixed to the values
obtained for the HETGS data (hereafter model G).
We also used the same continuum components (a power law and a
black-body), but we allow these parameters to vary freely in the fit,
as the continuum level and spectral shape might be different (in
addition, the LETGS is more sensitive to low energy photons and so
these data allows for a better determination of the black-body
temperature). In model G we also let free to vary the parameters for
the emission lines detected in the HETGs spectra (except for the Fe
k$\alpha$ and \sixiii\ lines that lie outside the spectral range of the LETG). We
further added three additional emission lines,  (located in wavelengths
larger than 25 \AA, the lower bound of the MEG data), corresponding to
\nv$\lambda 29.502$, \nvi$\lambda 29.029$, and \cv$\lambda 41.429$).

An acceptable fit is obtained to the LETGS spectrum with this simple method ($\chi^2/d.o.f.= 2013/1618$), implying that the HETGS spectra alone can constrain in a reliable way the properties of the ionized absorber. The best-fit continuum parameters for the LETGS spectrum are collated in Table \ref{tab:contMH}.
Table \ref{tab:EL} presents the parameters for the emission lines.

The best-fit model over the data is presented in Figure \ref{fig:leg}
(see also Table  \ref{tab:models}). This figure show that most
absorption lines observed in the HETG data  are well reproduced by the
model in the LETGS spectrum. It also shows
that model G fits well more than 12 low ionization absorption lines in
the region  (26-49)~\AA. These lines correspond to transitions of ions
such as \cv, \nvvi,~and \sxii. In the UTA region there are discrepancies between data and model, but as for the HETGS data, they are produced likely due to the use of the abbreviated data to model this feature.

We note that some low ionization lines show residuals consistent with
an outflow velocity of -1040 km s$^{-1}$ as  shown in the Figure
\ref{fig:lin5to}. This may indicate the presence of a HV-LIP (see \S
\ref{D}). However, attempting to model these residuals with an additional
absorbing component does not improve the fit in a statistical way.


\section{ Discussion \label{D}}

The best fit model of the WA for the Seyfert 1 galaxy  NGC 5548 requires two
outflow velocity systems, each with two different ionization phases
(\S  \ref{model}). Figure \ref{fig:lines2c} shows that the profiles of
the absorption lines (produced by ions such as \nex, \fexviii,
\oviii,~and \mgxi) are fit better with two narrow line profiles than with a
single broad component. Each of these two velocity components requires absorption by gas with
very different ionization level. The LV component consists of gas with two
different ionization states (the LV-LIP and the LV-HIP). The HV
component requires also two absorbers with different ionization  (the
HV-SHIP and the HV-HIP). These four absorbers, divided into two
velocity systems, can reproduce all the spectral features observed in
the spectra (\feviixxiv, \oviviii, \neixx, \mgixxii, \sixiixiv, \cvvi, \sixxii,~and \arxixii).
We note that the HV system might also have an additional, low
ionization component (HV-LIP). Figure \ref{fig:lin5to} shows a double
line profile for several low ionization level lines,
suggesting  a LIP with higher outflow velocity (-1041 km s$^{-1}$).
While the inclusion of this component is not statistically required
by the data, its physical properties appear to be well constrained (as can be observed in in Figure \ref{fig:5aregp}, where
the 1, 2, and 3$\sigma$ confidence regions for the ionization parameter
vs. the  column density of this component are presented). Further
evidence for the presence of this component arises from the fact that
UV system 1 can be associated with the HV absorber (see \S \ref{C}),
and the HV-SHIP and HV-HIP have a very hgh ionization state to produce
such absorption ((see Tables \ref{tab:ionden} and \ref{tab:ionden2}).   
We conclude that the data clearly shows a shift to higher ionization
material at higher velocity in NGC 5548 WA, with the LIP becoming weak
and the SHIP dominating in the HV system. On the other hand, the LV
system does not show the presence of the SHIP.

The separation between components is a new and
important feature in X-ray spectra of ionized absorbers, and could
be achieved because of the large difference in velocity between the two systems ($\sim 500-600$  km s$^{-1}$), as well as the different dominant
ionization levels between them (see also Steenbrugge et al. 2005).

\subsection{ UV Counterparts \label{C}}

In the study of the narrow \ovi, \nv,~and \civ\ absorption lines in the UV region, Crenshaw et al. (2003) finds five separate kinematic components in $\hst$ -STIS and $\fuse$ spectra. Our analysis reports two kinematic components (HV and LV systems). The HV system shows similar outflow velocity to the UV absorbing component 1 ($-1041$ km s$^{-1}$). According to Crenshaw et al. (2003)b, the gas producing absorbing component 1 should produce only a weak X-ray WA, which is consistent with the marginal detection of the HV-LIP in the \chandra\ data (\S\ \ref{D}).

There is also a correspondence between the X-ray LV system and UV
absorbing components 2 and 3 (with outflow velocities $-667$ and
$-530$ km s$^{-1}$, respectively). Crenshaw et al. (2003)b and
Crenshaw et al. 2009 suggest that at least component 3 (and possibly
component 2) should produce absorption in the X-ray range.
This is consistent with the detection of the LV-LIP.
There is no X-ray detection of UV absorbing systems 4 and 5 (with velocities $-336$ and $-166$ km s$^{-1}$ respectively), although we cannot rule out a possible contribution from component 4 to the LV-LIP, given the uncertainty in the outflow velocity of the X-ray systems. According to the analysis of the UV data by Crenshaw et al., system 5 has the lowest column density (logN$_H$=19.27).
 Krongold et al. (2003) discuss how the difference in sensitivity of the
current X-ray and UV detectors may affect the detection or non-detection of a given
UV component in the X-rays. Their Figure  11 shows the observational
lower limits for the equivalent widths of the \civ$\lambda\lambda1548.2, 1551$, \ovi$\lambda\lambda 1038, 1032$, \ovii$\lambda21.602$, \oviii$\lambda18.968$ lines detectable in the UV and X-ray bands, as a function of the ionization parameter and 
H equivalent column density. Thus, given their ionization state and column density, systems 4 and 5 are consistent with not being detected in the X-rays, given the limited sensitivity of the X-ray observations.

While a kinematic connection between the components is evident, an  
important question is how the predicted/measured column densities for
the different analyses compare between them.   
The ionic column densities predicted by our model are now listed in Table \ref{tab:ionden}
and \ref{tab:ionden2} for the LV and HV systems, respectively. In these Tables, we
also list the column densities predicted (or measured) by Crenshaw et
al. (2009), based on UV data, and by Steenbrugge et al. (2005), based
on the LETG 2002 data (their model B, see \S \ref{CPS}).
The column densities for ions CIV, NV y OVI between our predicted values and
those measured by Crenshaw et al. (2009) in the UV differ by factors of ~ 16, 4 and 1.2,
respectively. We note that OVI (where no diference is found) is detected in the X-ray analysis as
well. The differences between the UV and X-ray columns could be
produced because of 3 different factors: 1) The blending of UV components 2
and 3 in the X-rays (and maybe an additional contribution from UV
component in our LV-LIP model. 2) The different SEDs used in the 2
analysis: The SED used by Crenshaw et al. includes much more photons
in the far (unobservable) UV, and 3) A detailed analysis on the
dependence of the covering factor on velocity could not be performed
in the analysis of the UV line profile. This could underestimate the
columns as shown by Arav et al. 2003.
Taking into account these effects, we consider that the predictions of
our model for the columns measured in the UV are very reasonable, and
conclude that the LV-LIP component can produce the absorption features
produced by UV component 3 (and maybe 2 and 4).
However,  we note that the values predicted by Crenshaw et al. for the
high ionization charge states differ from those measured in this analysis. The
reason for this discrepancy, is that these ions are not observable in
the UV, and only three ions (CIV, NV y OVI) were used in the UV model.

With these considerations, and given the strong coincidence in
kinematics between the UV and X-ray outflows, we conclude  that a
common origin is not only possible, but quite likely. Such connection was originally suggested by Mathur et al. (1994) for NGC 5548, and has also been suggested for several other objects (NGC 985- Krongold et al. 2005; NGC 3783-Gabel et al. 2003, Krongold et al. 2003, Netzer et al. 2003; NGC 7469- Blustin et al. 2003, Kriss et al. 2003). Thus, these studies on the WA may represent a general feature of the structure for
quasars, as proposed by Elvis (2000).

\subsection{Two Multi-phase Absorbing Outflows} \label{S-Curv}

The presence of two different absorbing components with different temperature but
similar outflow velocity suggests that the absorber may be formed by a
multi-phase medium (e.g. Elvis, 2000; Krongold et al. 2003). Given the two
velocity systems in NGC 5548, we could be seeing the presence of two
different outflows (with different outflow velocity), each being a
multi-phase medium.

This is further supported by the pressure equilibrium of the two absorption components in each
velocity system. In Figure \ref{fig:sc1} we show the thermal stability curve of the gas (also known as the
"S-curve", Krolik et al. 1981) for the SED used in our analysis (\S 3.2.1; Fig. \ref{fig:sed}). The S-curve
marks the points of thermal equilibrium in the T vs. U/T plane,  where T is the photoionization
equilibrium temperature of the gas, and U/T is inversely proportional to the gas pressure (U/T $\propto P_{rad}/P_{gas}$ ).

In Figure  \ref{fig:sc1}a, the position on the S-curve of the two LV absorbing
components is shown. The LV-HIP and the LV-LIP have quite different temperatures
but, within the errors,  are consistent with a single value of the U/T, and so can be in pressure equilibrium.
Figure \ref{fig:sc1}b shows the HV-SHIP and HV-HIP on the S-curve. Again, the pressure equilibrium
between the two components is allowed. Note that, given the possible
range of ionization parameters of the tentative HV-LIP (Figure
\ref{fig:5aregp}), this component, if indeed present (as suggested by
our results and by those by Crenshaw et al. 2009),  could also be in pressure equilibrium
with the higher ionization ones that form this system. 
Thus, both the HV and the LV systems
are formed by a two-phase medium  (or maybe three-phase for the HV system), with one component pressure confining the others (see Krongold et al. 2005a for a more detailed explanation). 

NGC 5548 thus joins the group of AGNs for which pressure equilibrium
applies: NGC 3783 (Krongold et al. 2003, Netzer et al. 2003), NGC 985
(Krongold et al. 2005, 2008), NGC 4051 (Krongold et al. 2007), Mrk 279
(Fields et al. 2007), UGC 11763 (Cardaci et al. 2009).  No other well
studied WA in a Seyfert galaxy is inconsistent with the description of
a multi-phase medium. We note however, that we will further test this
idea for the WA over a large sample of objects (Andrade-Velazquez et
al. in preparation). NGC 5548 is the first AGN in which a two-phase WA
medium is likely for two velocity systems. The presence of a
multi-phase medium obviously requires that the different ionization
components that form it lie at the same distance from the ionizing
continuum.  We do not have the means to test this
requirement (though see Krongold et al. 2010). However, for NGC 4051, Krongold et al. (2007) 
measure the distance of the two absorbing components and found that
they were indeed consistent with being co-located, giving further plausibility to
multi-phase absorbing winds.

The SHIP of the HV system and the HIP of both the LV and HV systems
lie on unstable regions of the S-curve, implying a possible
inconsistency with the multi-phase scenario. However, the shape of
this curve depends on the exact shape of the SED of the source, which
has major uncertainties, particularly in the extreme UV region
(Haro-Corzo et al. 2007).  The chemical composition of the absorbing
gas has an important effect on this curve also (see Komossa \& Mathur,
2001, Fields et al. 2007). In addition, recent studies of the S-curve
with new and more reliable dielectronic recombination rate
coefficients (Badnell et al. 2006) show that there is a larger
probability of having a thermally stable WA at 10$^{5}$ K, (Crenshaw
et al. (2009) obtain a T $\sim$ 1$\times10^5$ for the UV component
3), and then a greater plausibility for its multiphase nature
(Chakravorty et al. 2008).



\subsection{Comparison with Previous Works} \label{CPS}

Steenbrugge et al. (2005) analyzed three \chandra\ grating observations of NGC 5548
(those carried out on 2002, Table 1) with a total
exposure time of 494 ks. They presented a model showing that a continuous distribution of
column densities and ionization parameters could fit the data. They
further concluded that the data was not consistent
with discrete phases of material in pressure equilibrium.

This conclusion is in contradiction to the results presented here, making a detailed
comparison of both analysis necessary. In their model C,
Steenbrugge et al. (2005) explored a model similar to the one presented
here. They used a self-consistent model that takes advantage of
photoionization balance calculations to make a global fit to the
data. They concluded that following this approach, and fitting Fe
separately, a good fit could be obtained, but constraining only two
absorbing components. They further tested the
presence of discrete components by adjusting equivalent H column
densities and ionization parameters to the absorbing components of
their model B, where ionic column densities were measured on an ion by
ion fit. In this case they find that at least five different
components with different ionization state were required. This approach led
them to conclude that discrete components were not a good
representation of the data.

The most fundamental difference between the analysis carried out by
Steenbrugge et al. (2005) and the present analysis is that their models B and C used only a single
velocity outflow with a large turbulent broadening to fit the data. However, as they report in their paper, and as discussed here, the data are better fitted with two different velocity components, the HV system and the LV system. Furthermore, as shown here (and as also noted by Steenbrugge et al.) the degree of ionization is different for the different velocity components,  as the HV system becomes dominant for high ionization states. This led them to conclude that more and more ionization components were required to fit the data and thus a structure of discrete clumps of gas was unlikely. However, here we have shown that, when each velocity component is modelled independently, both can be well described by two ionization components, making the presence of discrete components plausible and likely. Furthermore, the addition of extra components does not improve the fit. So, given the difference between number of free parameters from the different models,  we can claim by Occam's razor that our global approach is preferable. Table \ref{tab:ionden} and \ref{tab:ionden2} show the column densities for our model and the Steenbrugge et al. model B. The differences found between our analysis and that by Steenbrugge et al.  (that analysis shows sistematically larger columns than ours) are likely produced because only one velocity component was used by  Steenbrugge et al. to model the line profile produced by the two velocity systems. A detailed fitting including the line profile of each velocity system produces more reliable column densities.

Steenbrugge et al. (2005) further tested whether their five discrete components
could be in pressure balance, and they concluded that they could
not. In fact, their S-curve (their Fig.10) has no region
where the slope is negative, making it impossible for the gas to be in
pressure balance. We stress again that the shape of the S-curve
depends strongly on the metallicity of the absorbing gas, and the
shape of the SED used to produce it (in this
work we use the best multiwavelength data available extracted from
NED). Given this, and the new results by  Chakravorty et al. (2008), we
conclude that the possibility that the different phase are in pressure
equilibrium cannot be discarded.  Here we have shown that when the
velocity systems are modelled independently, each one is consistent
with pressure balance. We further note that Chelouche (2008) also find
that the WA in NGC 5548 can be described by a multiphase medium. The author
models the ionized flow assuming a thermal wind. While his model is
more physically driven than ours, it  does not take into account the
two different velocity systems revealed by the data.
However, given the similar results, the idea of a multi-phase medium
absorber is reinforced.

\subsection{Possible Geometry of the Ionized Outflow}

The presence of two different velocity systems, each formed by gas
with very different ionization state, can give a clue about the structure
of the flow. In a continuous {\bf radial} range of ionization structures, the wind must extend in our line of sight for parsecs (or more),  forming a large scale outflow along our line of sight. We find difficult to reconcile such a distribution of  material with the two velocity systems found
in NGC 5548, even if inhomogeneities in the flow (like those suggested by Ram\'irez et al. 2005) are invoked. The presence of the two velocity components, then, suggests that we are not seeing the wind in a radial direction.

On the other hand, the presence of the different kinematic components
can be naturally explained if we are looking at an outflow in a
transverse direction, which is consistent with the detection of
transverse motion in the UV absorbers (Marthur et al. 1995, Arav et
al. 2002, Crenshaw et al. 2003). In this case, an important component of the velocity
(and acceleration) of the wind could take place in the direction perpendicular to our line of sight.
Such a configuration for the wind is further supported by the finding
that the different phases
observed at each velocity form a single multi-phase outflow. 

If the wind arises from the inner accretion disk (Konigl \& Kartje,
1994, Elvis et al. 2000,  and Krongold et al 2007), and flows in a
transverse direction with respect to our line of sight, the two
velocity components could be simply explained as forming in different
regions of the disk (Elvis, 2000). The two velocity components in
NGC 5548 are consistent with originate in the accretion disk, as
discussed by Krongold et al. (2010).  
Furthermore, Crenshaw et al. (2009) suggest that UV component 3
(clearly associated to the LV-LIP), should be accelarated
by magnetocentrifugal forces, and thus should form at an accretion-disk
scale (Bottorff et al. 2000). 
We note that an accretion-disk wind does not imply a small scale outflow. The wind
could still extend for large scales, but our line of sight would cross
it only at a given distance from the central ionizing source. 

\section{Conclusions \label{Sum}}

We find that the WA of NGC 5548 can be
modelled in a simple way with two different velocity systems
(V$_{out}$ = -1110 km s$^{-1}$ and
V$_{out}= -490\pm49$ km s$^{-1}$) each with two
phases of gas in pressure equilibrium. This results strongly
suggests that the structure of the absorber in
NGC 5548 could consist of two different outflows
(with different outflow velocity), each formed by a multi-phase medium
in pressure balance. The ionized absorber in
NGC 5548 now fits a pattern established for the other well studied
systems. There are no remaining cases in which a multi-phase medium is
inconsistent with the observations.

The presence of two velocity systems, each formed by a multi-phase
medium suggests that we are looking at the flows in a
transverse direction with respect to our line of sight (as also suggested
by the UV data).

The velocity of the two X-ray absorbing systems is in agreement with
components  1, 2, and 3 found in the UV band (Crenshaw et al. 2003b).
The X-ray data are further consistent with the non detection of UV
components 4 and 5, given the limited sensitivity and resolution of
the current X-ray (compared to those in the UV). There is also a
reasonable agreement in the column densities of CIV, NV, and OVI
predicted by our X-ray analysis, and those measured in the UV. Thus, our models
elegantly solve the long-standing problem of apparent discrepancies
between the UV and X-ray absorbers, and give further
support to the idea that UV and X-ray absorbers are part of the
same outflow.

\acknowledgements
This work was supported by the UNAM
PAPIIT grant IN118905 and the CONACyT grant J-49594. A-VM acknowledges support from CONACyT scholarship. NSB acknowledges support from NASA to the Chandra X-ray Center through NAS8-03060. This work was supported by NASA grant NNX08AB81G.

\appendix
\section{\bf APPENDIX \label{app}}

In this appendix we have enumerated (from high to low energy) the 31 references for the 185
photometric data points from  NED used to build the
Spectral Energy Distribution.

\begin{enumerate}

\item    
Kaspi, Shai; Maoz, Dan; Netzer, Hagai; Peterson, Bradley M.; Vestergaard, Marianne; Jannuzi, Buell T. \ 2005, \apj, 629, 61K

\item    
McKernan, B.; Yaqoob, T.; Reynolds, C. S.
\ 2007, \mnras, 379, 1359M

\item
Jack W. Sulentic, Rumen Bachev, Paola Marziani, C. Alenka Negrete, and
Deborah Dultzin \ 2007, \apj, 666, 757S

\item    
Mu\~noz Mar\'in, V\'ictor M.; Gonz\'alez Delgado, Rosa M.; Schmitt, Henrique R.; Cid Fernandes, Roberto; P\'erez, Enrique; Storchi-Bergmann, Thaisa; Heckman, Tim; Leitherer, Claus
\ 2007, \aj, 134, 648M

\item    
Anderson, Kurt S. \ 1970, \apj, 162, 743A

\item    
McAlary, C. W.; McLaren, R. A.; McGonegal, R. J.; Maza, J. \ 1983, \apjs, 52, 341M

\item
de Vaucouleurs, G.; de Vaucouleurs, A.; Corwin, H. G., Jr.; Buta, R. J.; Paturel, G.; Fouque, P. \ 1991, RC3.9, C.

\item    
Koulouridis, E.; Plionis, M.; Chavushyan, V.; Dultzin-Hacyan, D.; Krongold, Y.; Goudis, C.
\ 2006, \apj, 639, 37K

\item    
Suganuma, Masahiro; Yoshii, Yuzuru; Kobayashi, Yukiyasu; Minezaki, Takeo; Enya, Keigo; Tomita, Hiroyuki; Aoki, Tsutomu; Koshida, Shintaro; Peterson, Bruce A.
\ 2006, \apj, 639, 46S

\item    
Zwicky, Fritz; Herzog, E. \ 1963, CGCG2.C...0000Z

\item    
Takamiya, M.; Kron, R. G.; Kron, G. E. \ 1995, \aj, 110, 1083T

\item
Petrosian, Artashes; McLean, Brian; Allen, Ronald J.; MacKenty, John W.
\ 2007, \apjs, 170, 33P

\item    
Bentz, Misty C.; Denney, Kelly D.; Cackett, Edward M.; Dietrich, Matthias; Fogel, Jeffrey K. J.; Ghosh, Himel; Horne, Keith D.; Kuehn, Charles; Minezaki, Takeo; Onken, Christopher A.; Peterson, Bradley M.; Pogge, Richard W.; Pronik, Vladimir I.; Richstone, Douglas O.; Sergeev, Sergey G.; Vestergaard, Marianne; Walker, Matthew G.; Yoshii, Yuzuru \ 2007, \apj, 662, 205B

\item
Bentz, Misty C.; Peterson, Bradley M.; Pogge, Richard W.; Vestergaard, Marianne; Onken, Christopher A. \ 2006, \apj, 644, 133B

\item
Lebofsky, M. J.; Rieke, G. H. \ 1980, Natur, 284, 410L

\item    
Wisniewski, W. Z.; Kleinmann, D. E. \ 1968, \aj, 73, 866W

\item    
Spinoglio, Luigi; Malkan, Matthew A.; Rush, Brian; Carrasco, Luis; Recillas-Cruz, Elsa
\ 1995, \apj, 453, 616S

\item    
Balzano, V. A.; Weedman, D. W. \ 1981, \apj 243, 756B

\item
McAlary, C. W.; McLaren, R. A.; Crabtree, D. R. \ 1979, \apj, 234, 471M

\item    
Rieke, G. H. \ 1978, \apj, 226, 550R

\item    
Pacholczyk, A. G.; Weymann, R. J. \ 1968, \aj, 73, 870P

\item    
Penston, M. V.; Penston, M. J.; Selmes, R. A.; Becklin, E. E.; Neugebauer, G.
\ 1974, \ mnras, 169, 357P

\item  Rodr\'guez-Ardila, A.; Riffel, R.; Pastoriza, M. G. \ 2005, \mnras, 364, 1041R

\item  de Vaucouleurs, Antoinette; Longo, Giuseppe \ 1988, VIrPh. C.

\item Stein, W. A.; Weedman, D. W. \ 1976, \apj, 205, 44S

\item  
Maiolino, R.; Shemmer, O.; Imanishi, M.; Netzer, H.; Oliva, E.; Lutz, D.; Sturm, E
\ 2007, \aap, 468, 979M

\item Kleinmann, D. E.; Low, F. J. \ 1970, \apj, 161L, 203K

\item   Rieke, G. H.; Low, F. J. \ 1972, \apj, 176L, 95R

\item  Gorjian, V.; Werner, M. W.; Jarrett, T. H.; Cole, D. M.; Ressler, M. E.
\  2004, \apj, 605, 156G

\item    
Moshir, M.; et al. \ 1990, IRASF, C, 0M

\item   Gorjian, V.; Cleary, K.; Werner, M. W.; Lawrence, C. R. \ 2007, \apj, 655L, 73G

\end{enumerate}


\begin{deluxetable}{lcccc}
\tablecolumns{5} \tablewidth{0pc} \tablecaption{Observations Log of NGC 5548.
\label{tab:obs}}
\tablehead{ \multicolumn{5}{c}{HETG/ACIS-S}}\startdata

Observation & UT Start Date & Exposure$^a$  & Flux$_{0.47-1.7 keV}^b$ & S/N(22\AA) \\
(2)837 &  2000-02-05 15:37 &  82.3 &  1.14&6.63\\
(3)3046  & 2002-01-16 06:12 & 153.9  & 1.96&4.69\\
TOTAL&&236.2&& 8.12 \\
\cutinhead{LETG/HRC-S}
(1)330 & 1999-12-11 22:51  &  85.98  &  2.99&34.27\\
(4)3045 &  2002-01-18 15:57&  169.68 & 1.38&35.71\\
(5)3383 & 2002-01-21 07:33 &  171.02 & 2.19&33.32\\
(6)5598  & 2005-04-15 05:18 &  116.43 & 0.56&14.45\\
(7)6268 & 2005-04-18 00:31 &  25.19  & 0.46&9.38\\
TOTAL&&568.3&&62.09 \\
\enddata
\tablenotetext{a}{ In ks}
\tablenotetext{b}{ In 10$^{-11}$erg s$^{-1}cm^{-2}$}
\end{deluxetable}

\begin{deluxetable}{llllllccccc}
\tablecolumns{11} \tablewidth{0pt}
\rotate
\tablecaption{Models used to fit the NGC 5548 spectra. \label{tab:models}}
\tabletypesize{\scriptsize}
\tablehead{\colhead{\bf M} &  \multicolumn{2}{c}{\bf Power Law}
&\multicolumn{2}{c}{\bf Black
Body} &  \colhead{\bf EL$^d$}&\multicolumn{4}{c}{\bf WA* } & \colhead{\bf Goodness} \\
& \colhead{$\Gamma$}  & \colhead{A$_{\rm pwlw}$\tablenotemark{a}} &
\colhead{kT (keV)}  &
\colhead{A$_{\rm BB}$\tablenotemark{b}} &
& \colhead{SHIP$_{HV}$} &  \colhead{HIP$_{HV}$} & \colhead{HIP$_{LV}$} & \colhead{LIP$_{LV}$} &  \colhead{$\chi^2$, dof, $\frac{\chi^2}{dof}$}}
\startdata
A  &    1.52$\pm0.006$  &  4.87$\pm0.02\times10^{-3}$ & 0.012$\pm$\tablenotemark{c} & 1.00$\times10^{-9}\pm$\tablenotemark{c} & 4 & -- & --& - & - -  & 3813, 4112, 0.93 \\

B &   1.560$\pm0.007$  &  5.20$\pm0.03\times10^{-3} $  & 0.085$\pm0.005$ &  6.98$\pm1.36\times10^{-5}$ & 4  & --&$\surd$ & - - & -  & 3053, 4108, 0.74  \\

C & 1.568$\pm$0.007     &  5.00$\pm0.03\times10^{-3}$   &  0.083$\pm$0.005  &7.28$\pm1.54\times10^{-5}$ & 4 &   $\surd$  &$\surd$  & - - & -    & 2987, 4104,
0.73 \\

D &  1.595$\pm0.008 $     &  5.00$\pm0.05\times10^{-3}$  &  0.099$\pm$0.006   & 5.65$\pm1.07\times10^{-5}$ & 4  &$\surd$ &$\surd$  & $\surd$ & -  & 2946, 4100,
0.72 \\

E & 1.597$\pm0.009$  & 5.40$\pm0.05\times10^{-3}$  & 0.110$\pm{0.007}$ & 5.17$\pm0.85\times10^{-5}$ & 4 &$\surd$  & $\surd$ & $\surd$&$\surd$  & 2904, 4096, 0.71 \\

F \tablenotemark{e} & 1.598$\pm0.010$  &  5.46$\pm0.05\times10^{-3}$  & 0.116$\pm{0.009}$ & 4.51$\pm0.75\times10^{-5}$ & 8 &$\surd$  & $\surd$ & $\surd$&$\surd$  &  2885, 4102, 0.70 \\

G  &  1.778$\pm0.02$  &  7.84$\pm0.05\times10^{-3}$  &  0.083$\pm{0.002}$ & 5.93$\pm0.02\times10^{-5}$ &  7 &$\surd$  & $\surd$ & $\surd$&$\surd$  &  2013, 1618, 1.24 \\

\enddata
\tablenotetext{a}{In units of $10^{-3}$ ph s$^{-1}$ cm$^{-2}$ keV$^{-1}$
at 1 keV}

\tablenotetext{b}{In units of $10^{-5}$  $L_{39}/D_{10}^2$,
where $L_{39}$ is the source luminosity in units of $10^{39}$ erg
s$^{-1}$  and $D_{10}$ is the distance to the source in units of
10 kpc}

\tablenotetext{c}{The uncertainties cannot be found due to the
presence of the absorber.}
\tablenotetext{d}{EL means Emission Lines.}
\tablenotetext{e}{Model E plus the other lines remains (see Table \ref{tab:EL}).}

\end{deluxetable}

\begin{deluxetable}{ccccc}
\tablecolumns{5} \tablewidth{0pc} \tablecaption{ Emission lines in the NGC 5548 spectrum\tablenotemark{a}.
\label{tab:EL}}
\tablehead{\colhead{ION} &  \colhead{ Rest Frame $\lambda$(\AA)\tablenotemark{b}} &  \colhead{Measure $\lambda$(\AA)} & \colhead{FWHM ($km s^{-1}$)}& \colhead{EW(m\AA)}}
\startdata
\cutinhead{HETG   SPECTRUM}
Fe K$_{\alpha}$ &  1.937 &  1.9368$^{+0.0086}_{-0.0070}$& 2644.4$\pm2469$ & 27.9 $^{+48.4}_{-13.3}$\\
\sixiii &6.648 & 6.739$^{+0.005}_{-0.005}$ &754$\pm938$ & 3.7 $^{+1.4}_{-1.4}$ \\
\neix & 13.699 &  13.678 $^{+0.005}_{-0.006}$ & 485$\pm593$ & 22.6$^{+31.7}_{-}$\\
\ovii  &21.607 &  21.597 $^{+0.018}_{ -0.018}$  &936$\pm636$ &  35.9$^{+19.6}_{-19.7}$\\
\ovii & 21.807 &  21.785 $^{+0.005}_{-0.007}$  & 928$\pm299$ & 73.2$^{+28.5}_{-25.5}$\\
\ovii & 22.101 &  22.075 $^{+0.007}_{-0.008}$  & 915$\pm311$ &  244.2$^{+54.8}_{-54.1}$ \\
\nvi &  23.024 &  23.035 $^{+0.014}_{ -0.304}$  & 327$\pm373$ & 26.2$^{+17.0}_{-17.0}$\\
\nvi &23.277 &  23.306$ ^{+0.010}_{-0.010}$  & 673$\pm398$ & 80.9$^{+29.3}_{-27.3}$\\

\cutinhead{LETG   SPECTRUM}

\neix & 13.699 &  13.677 $^{+0.016}_{-0}$ &60.7$\pm822$ &  5.9$^{+2.1}_{-2.2}$\\
\ovii &21.607 &  21.623 $^{+0.022}_{ -0.019}$  & 944$\pm513$ &  34.9$^{+9.8}_{-7.9}$\\
\ovii & 21.807 &  21.760 $^{+0.025}_{-0.013}$  & 935$\pm554$ & 42.8$^{+23.1}_{-8.0}$\\
\ovii & 22.101 &  22.075 $^{+0.004}_{-0.004}$  & 923$\pm159$ & 119.0$^{+10.1}_{-10.1}$ \\
\nvi &29.534 &29.502$ ^{+0.021}_{-0.016}$ &462$\pm884$ &23.9$^{+13.0}_{-9.5}$ \\
\nvi &29.081 & 29.029\tablenotemark{c}&10\tablenotemark{c} &11.5$^{+7.1}_{-0}$ \\
\cv & 41.472 &41.429\tablenotemark{c} &20$\pm452$ & 121.1$^{+38.8}_{-38.8}$\\
\enddata

\tablenotetext{a}{All line positions are reported with respect to the
rest frame of the host galaxy.
Given the uncertainties in the measurement, as well as the filling of
the absorption lines by the emission features, all lines are consistent with the rest frame of the host galaxy.}
\tablenotetext{b}{The line rest wavelengths were taken from the NIST Atomic Spectra Database (URL: http://physics.nist.gov/PhysRefData/ASD/index.html)}
\tablenotetext{c} {The uncertainties cannot be found.}
\end{deluxetable}

\begin{deluxetable}{cccc}
\tablecolumns {3} \tablewidth{0pc} \tablecaption{The best continuum parameters. \label{tab:contMH}}

\tablehead{ \multicolumn{3}{c}{Power Law}  } \startdata
Photon Index ($\Gamma$) & Normalization\tablenotemark{a} & $N_{HGal}$ (cm$^{-2}$)\\
HETGS \   1.59$\pm 0.010$ & 54$\pm 0.5$& $1.83\times10^{20}$ \\
LETGS  \   1.77$\pm 0.02$  &78$\pm 0.5$ & " " \\

\cutinhead{Thermal component}

kT (keV) & Normalization\tablenotemark{b} & \nodata \\

HETG \   0.11$\pm 0.009$ & 0.45$\pm 0.08$ & \nodata \\
LETG  \   0.083$\pm 0.0016$ &  0.59$\pm 0.002$ & \nodata \\

\enddata
\tablenotetext{a}{In $10^{-4}$ photons keV$^{-1}$ cm$^{-2}$ s$^{-1}$ at 1
keV.} \tablenotetext{b}{In $10^{-4}$ $L_{39}/D_{10}^2$, where $L_{39}$ is
the source luminosity in units of $10^{39}$ erg s$^{-1}$ and
$D_{10}$ is the distance to the source in units of 10 kpc. }
\end{deluxetable}

\begin{deluxetable}{ccccc}
\tablecolumns{5} \tablewidth{0pc} \tablecaption{ Models B-D Parameters for the Ionized Absorber \label{tab:comp}}
\tablehead{\colhead{COMPONENT} &  \colhead{ logU} &  \colhead{logN$_H$(cm$^{-2}$)} & \colhead{Vel$_{out}$ (km s$^{-1}$)}& \colhead{Vel$_{turb}$ (km s$^{-1}$)}}
\startdata
B1& 0.75$\pm$0.02 & 21.58$\pm$0.03& -740$\pm$150& 536$\pm$35\\
C1& 1.06$\pm$0.10& 21.59$\pm$0.06&-1120$\pm$150& 117$\pm$26\\
C2&0.63$\pm$0.02 &21.28$\pm$ 0.04&-450$\pm$150&241$\pm$37\\
D1&1.14$\pm$0.02&21.71$\pm$0.08&-1120$\pm$150&113$\pm$42\\
D2&0.67$\pm$0.02&21.30$\pm$0.04&-450$\pm$150& 275$\pm$117\\
D3&-0.49$\pm$0.08&20.74$\pm$0.10&-590$\pm$150&105$\pm$ 47\\
\enddata
\end{deluxetable}

\begin{deluxetable}{lcccc}
\tablecolumns{5} \tablewidth{0pc} \tablecaption{Final Best Fit Ionized Absorber
Parameters (Model E and F).
\label{tab:2abHV}}

\tablehead{\colhead{Parameter} & \multicolumn{2}{c}{V$_{hi}$=$-1110$ km s$^{-1}$}&\multicolumn{2}{c}{V$_{lo}$=$-490$ km s$^{-1}$}\\& {SHIP} &
\colhead{HIP}& {HIP} &{LIP}} \startdata
Log U$^a$ &  1.23$\pm0.06$ & 0.67$\pm0.03$&  0.67$\pm0.03$ & -0.49$\pm0.09$ \\
Log N$_{H}$ (cm$^{-2}$)$^a$ &  21.73$\pm0.12$ & 21.03$\pm0.07$ &  21.26$\pm.04$ & 20.75$\pm0.10$\\
V$_{Turb}$ (km s$^{-1}$)& 175 $\pm67.0$ &  100 $\pm23.0$ & 177 $\pm39.0$& 105 $\pm29.0$\\
V$_{Out}$ (km s$^{-1}$)$^a$& -1040$\pm150$ & -1180$\pm150$& -400$\pm150$ & -590$\pm150$ \\
T (K)$^b$& $27.0\pm5.6$ & $5.8\pm1.0$ & $5.8\pm0.8$ & $0.35\pm0.04$\\
$[$Log T (K)$]$& 6.47$\pm0.12$ & 5.89$\pm0.06$ &5.72$\pm0.12$ & 4.54$\pm0.02$\\
Log T/U ($\propto$P$^c$)& 5.21$\pm0.17$ & 5.15$\pm0.15$& 5.04$\pm0.14$ & 5.06$\pm0.1$ \\
\enddata
\tablenotetext{a}{Free parameters of the model.}
\tablenotetext{b}{In units of 10$^5$ Kelvin. Derived from the column density and ionization
parameter, assuming photoionization equilibrium.}
\tablenotetext{c}{The pressure P$\propto$n$_e$T. Assuming that
both phases lie at the same distance from the central source
n$_e\propto$1/U, and P$\propto$T/U.}
\end{deluxetable}

\begin{deluxetable}{lccccc}
\tablecolumns{6} \tablewidth{0pc} \tablecaption{Ionic Column Densities
  for the LV system\label{tab:ionden}}
\tablehead{\colhead{ION} &  \colhead{ logN$_{LIP}$} &  \colhead{logN$_{HIP}$} &\colhead{logN$_{SUM}$} &  \colhead{ logN$_{iC}$}&\colhead{logN$_{iS}$}}
\startdata
\civ &  15.20 &  10.13 & 15.20 & 13.99     &  ------ \\
\cv &   \textbf{16.75}   & 	13.76   & 16.75  &   16.90 &    17.20     \\
\cvi &  \textbf{17.03}   & 16.03   &   17.07 & 18.10  &     17.60       \\
\nv &  \textbf{15.53}	 &   11.07  & 15.53  &  14.93   &  -------\\
\nvi &   \textbf{16.46} &  14.06 &  16.46    &    17.20   &   17.20    \\
\nvii &  \textbf{16.26}  &   15.86  & 16.41 &    18.0 &     17.50   \\
\ov &  	 \textbf{16.36}    &  11.11   & 16.36 & 	15.70	& 	16.80	\\
\ovi  &	 \textbf{  16.89}  &   	13.15 &  16.89 &   16.80   &    16.60 \\
\ovii &  \textbf{17.40}  &   \textbf{ 	15.68}  & 17.41 &   18.50 &     18.18   \\
\oviii &  \textbf{ 16.78}  &  \textbf{17.06}   & 17.24 &   18.70   &     18.53  \\
\neix &   \textbf{15.92} &   \textbf{16.10}    &   16.32 &  18.00  &    17.20 \\
\nex &  	14.70  &   \textbf{16.88}     &   16.88 & 17.60 &   17.90     \\
\mgxi &  15.04  &  	\textbf{16.20}   &  16.23 & 17.30 &     17.00 \\
\mgxii  & 	13.31    & \textbf{16.51}     &   16.51 & 16.50   &   17.30  \\
\siviii &  	15.46  &  12.06 &  15.46 &  16.70  &    16.20  \\
\siix &  15.83  &   	13.40   &   15.82 &  17.20  &    16.00    \\
\sixi &  	15.44   &  14.34   & 15.47  &   16.80  &     16.00  \\
\sixiii &  	13.81    &  \textbf{16.45}   & 16.45  &    16.40  &     17.00  \\
\sixiv & 	11.74   &  \textbf{16.39}     & 16.39  &     15.20 &     17.00  \\
\sxi & 15.35& 	14.99  &  15.51 & 16.90 &  16.00  \\
\sxii &  	14.84  &   15.63  &   15.70 &	16.70  &  16.20 \\
\fexvii &  	13.82 &  15.73 &   15.74 &  15.10 &    16.30  \\
\fexviii & 13.54  & \textbf{16.19}  & 16.19 &	------   & ------- \\

\enddata
\tablenotetext{a}{Columns 2 and 3 correspond to the
    ionic column densities predicted by the LV system of the best
    model in this work. 
Column 4 is the added contribution of the two ionization components
(LV-LIP and LV-HIP) of this system.  The predicted ionic column
densities by UV component 3 (Crenshaw et al. 2009) are located in
column 5. Column 6 shows the column densities reported by model B of Steenbrugge
et al. 2005. All columns are in units of cm$^{-2}$.}

\end{deluxetable}

\begin{deluxetable}{ccccc}
\tablecolumns{5} \tablewidth{0pc} \tablecaption{Ionic Column Densities
  for the HV system\label{tab:ionden2}}
\tablehead{\colhead{ION} &  \colhead{logN$_{HIP}$} & \colhead{logN$_{SHIP}$}&   \colhead{logN$_{SUM}$} &\colhead{logN$_{iS}$}}
\startdata
\civ &   10.36   &  9.04  & 10.38  &  -----   \\
\cv &     12.98 &  	12.74    & 13.18 &    17.20    \\
\cvi &      \textbf{16.26}   &  15.63   & 16.35 &    17.60   \\
\nv &     11.30  &  9.57    & 11.31 &    ------   \\
\nvi &     14.30 &  	12.96  & 14.32 &   17.20    \\
\nvii &     16.09    &  15.50   &  16.19 &    17.50     \\
\ov &    10.86 &   9.30 & 10.87 & 16.80	\\
\ovi  &	   13.37 &  11.30    &  13.37 &   16.60  \\
\ovii &    15.90  & 14.51 &   15.92 &  18.18   \\
\oviii &  \textbf{17.29}    &   \textbf{	16.76 }  & 17.40 &   18.53   \\
\neix &  	\textbf{15.86}   &   	14.68 &  15.89 &  17.20  \\
\nex &    \textbf{	16.65}     &  \textbf{16.55}  & 16.90 &  17.90    \\
\mgxi &  	\textbf{15.96}     &  15.10 &  16.02&  17.00  \\
\mgxii  & \textbf{16.28}  &  \textbf{16.44}	 & 16.67 &  17.30 \\
\siviii & 11.83 &  7.98   &  11.83 &  16.20     \\
\siix &  	13.16 &  7.98   &   13.16 &  16.00    \\
\sixi &  	14.10  &  11.85    &  14.10 &  16.00   \\
\sixiii &  \textbf{16.22} &  15.75    &   16.35 &   17.00  \\
\sixiv & \textbf{16.16}  &  \textbf{16.71}   & 16.82&  17.00 \\
\sxi & 14.17 &  9.71 & 14.17 & 16.00  \\
\sxii &   14.75   &  12.37  &   14.75 & 16.20 \\
\fexvii & 15.49 & 13.24&  15.49 & 16.30   \\
\fexviii & 	\textbf{15.96}  & 14.12 & 15.97  & ------- \\

\enddata
\tablenotetext{a}{Columns 2 and 3 correspond to the
    ionic column densities predicted by the HV system of the best
    model in this work. 
Column 4 is the added contribution of the two ionization components
(HV-HIP and HV-SHIP) of this system. Column 6 shows the column
densities reported by model B of Steenbrugge
et al. 2005. All columns are in units of cm$^{-2}$.}
\end{deluxetable}


\FloatBarrier{
\begin{figure}

\caption{Model F (red thick line; corresponding to model E plus 4 additional emission lines, see \S\ 3.4 ) plotted over the MEG spectra ([5-24.5] \AA) of NGC~5548. Two velocity components, each formed by two phases, are required to fit the data. The upper labels correspond to the absorption lines, while the lower labels mark the emission features.}\label{fig:meg}
\plotone{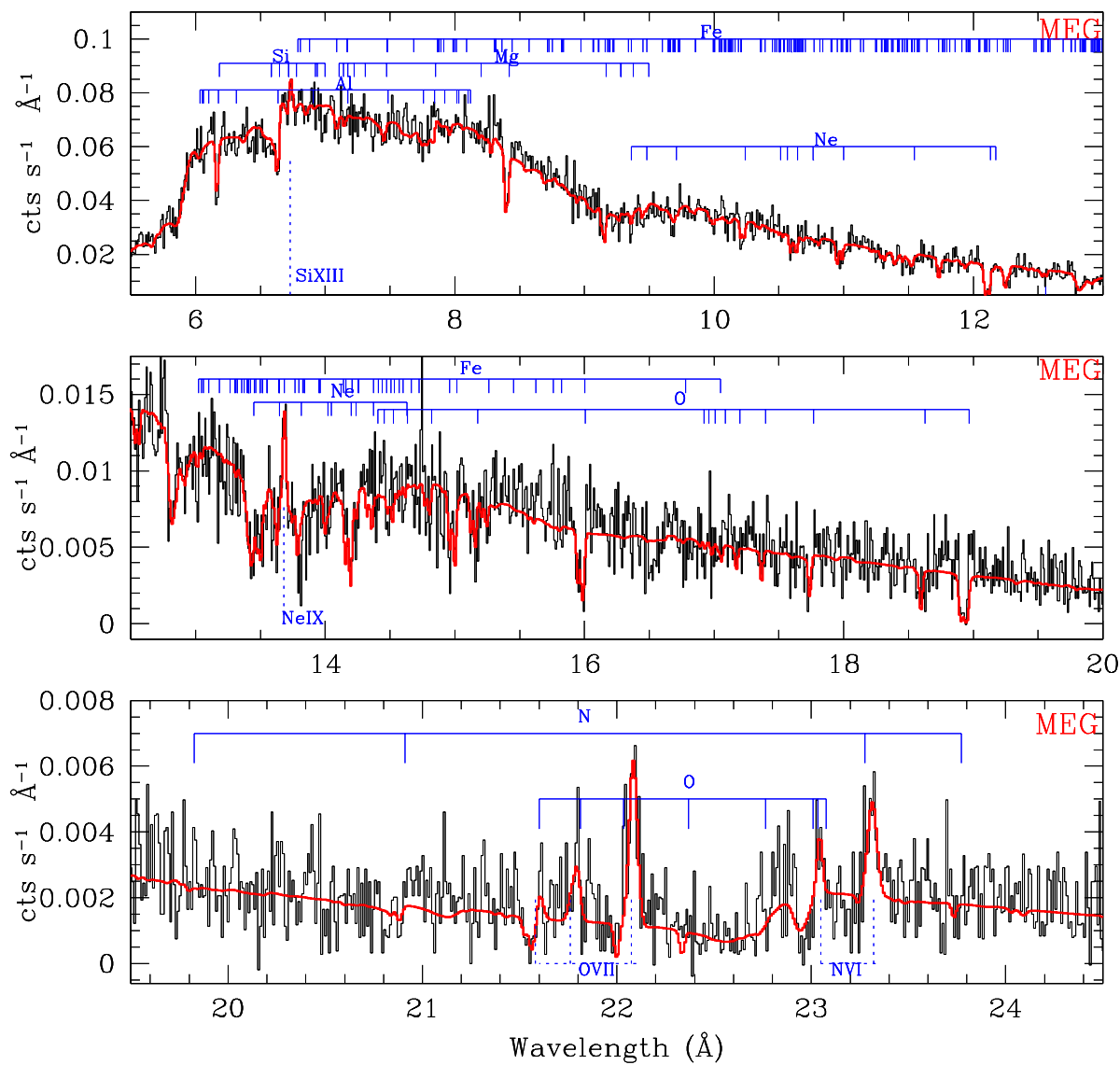}
\end{figure}

\begin{figure}
\caption{Model F (red thick line) plotted over the HEG spectra
([1.6-8.84]~\AA) of NGC~5548.}
\plotone{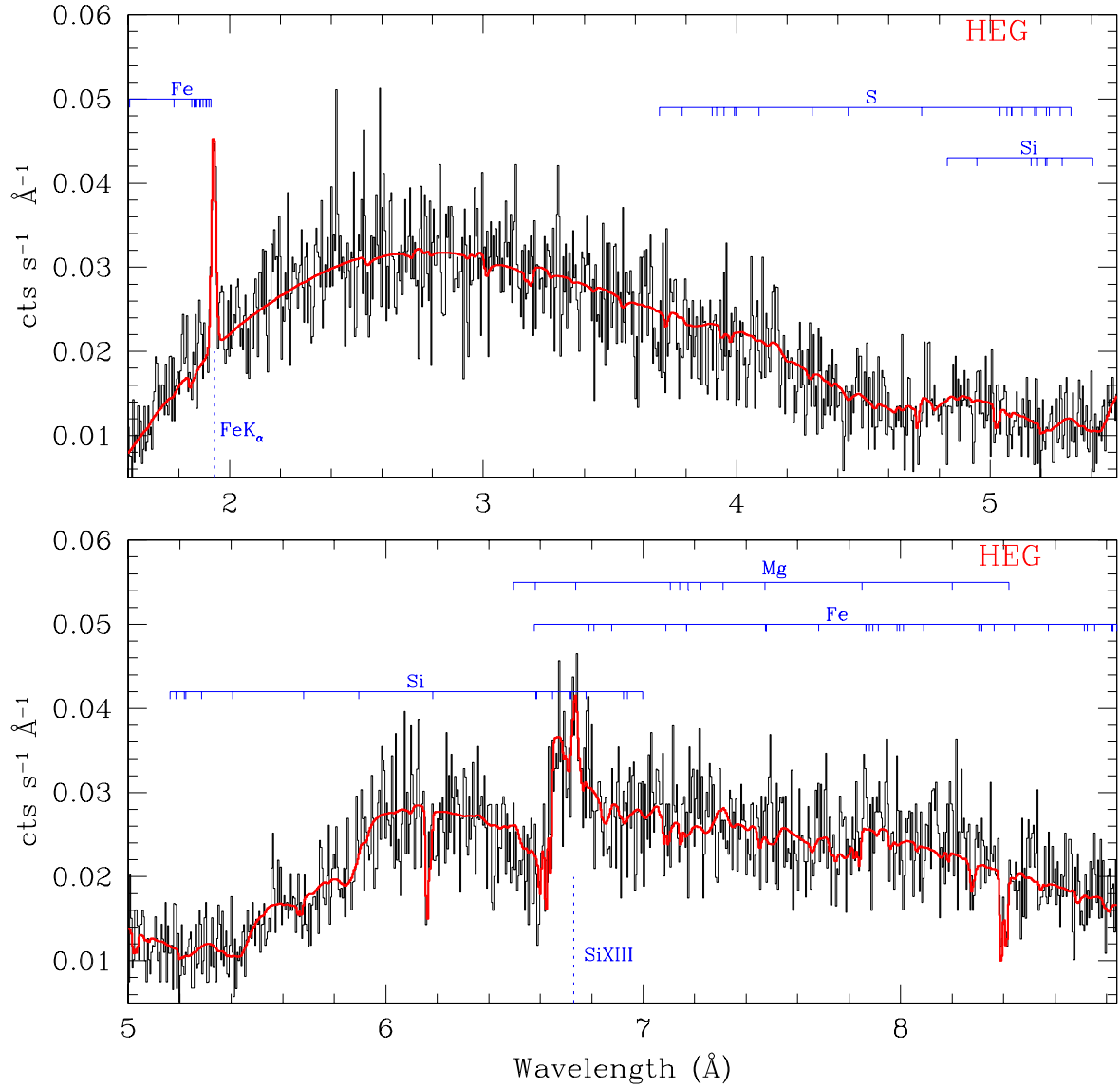}\label{fig:heg}
\end{figure}

\begin{figure}
\caption{MEG spectrum and model between  (5 and 24.5) \AA\ showing only the
contribution of the low outflow system. The left panel shows the contribution by the Higher Ionization Phase (HIP) and the right panel shows the Low Ionization Phase (LIP). }\label{fig:LV}
\plottwo{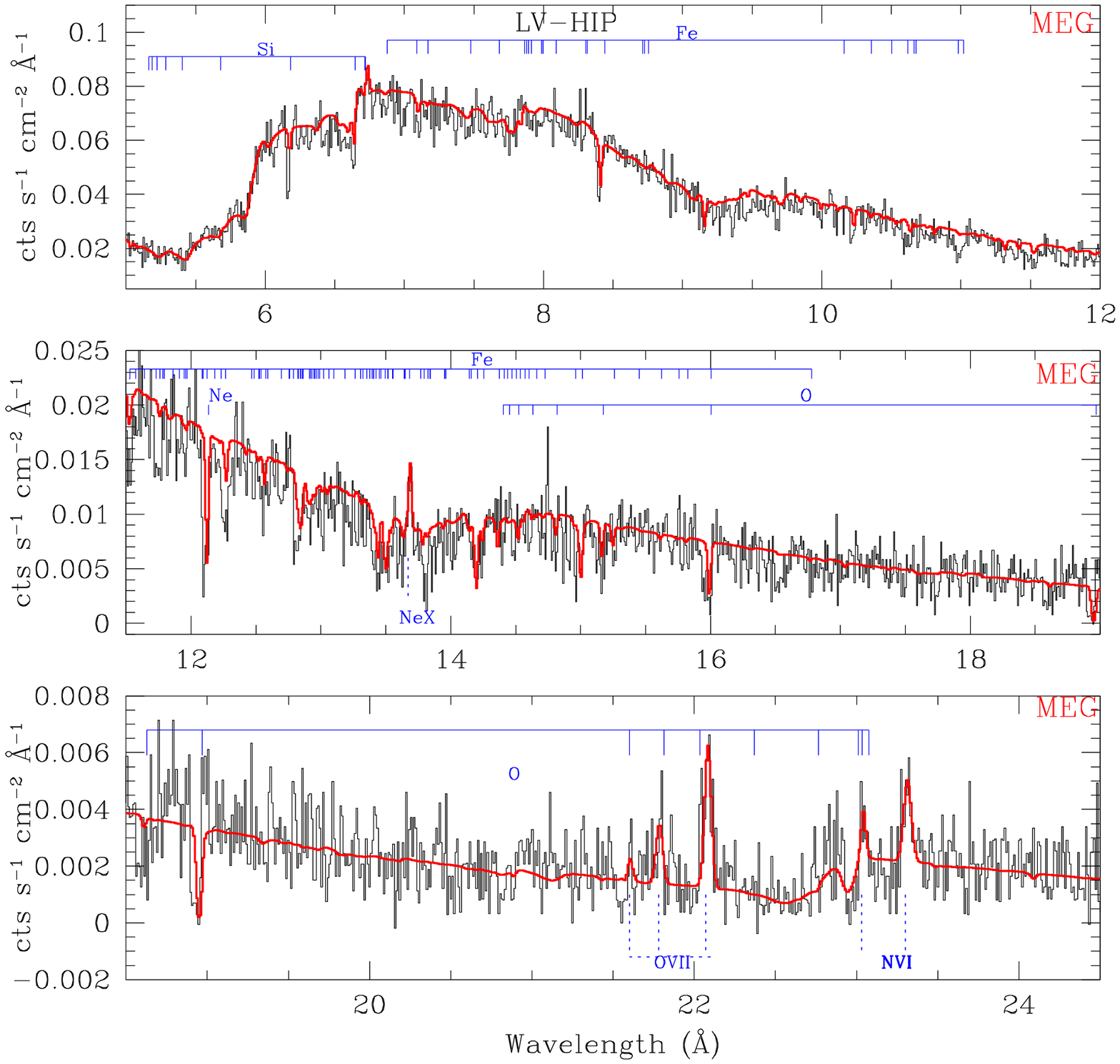}{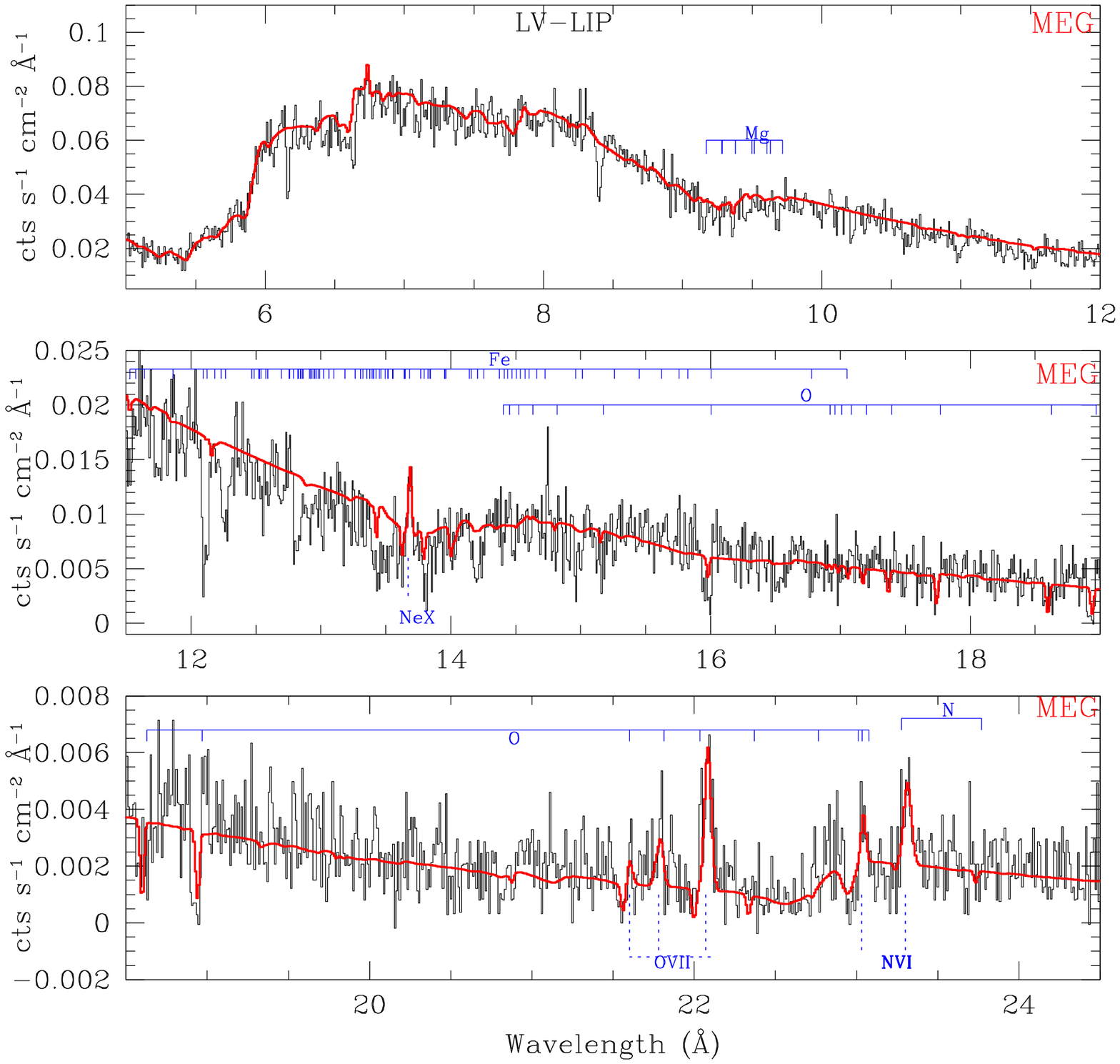}
\end{figure}

\begin{figure}
\caption{MEG spectrum and model between (5-24.5) \AA\ showing only the contribution of the HV-HIP,  and the right panel the HV-SHIP.}
\label{fig:HV}
\plottwo{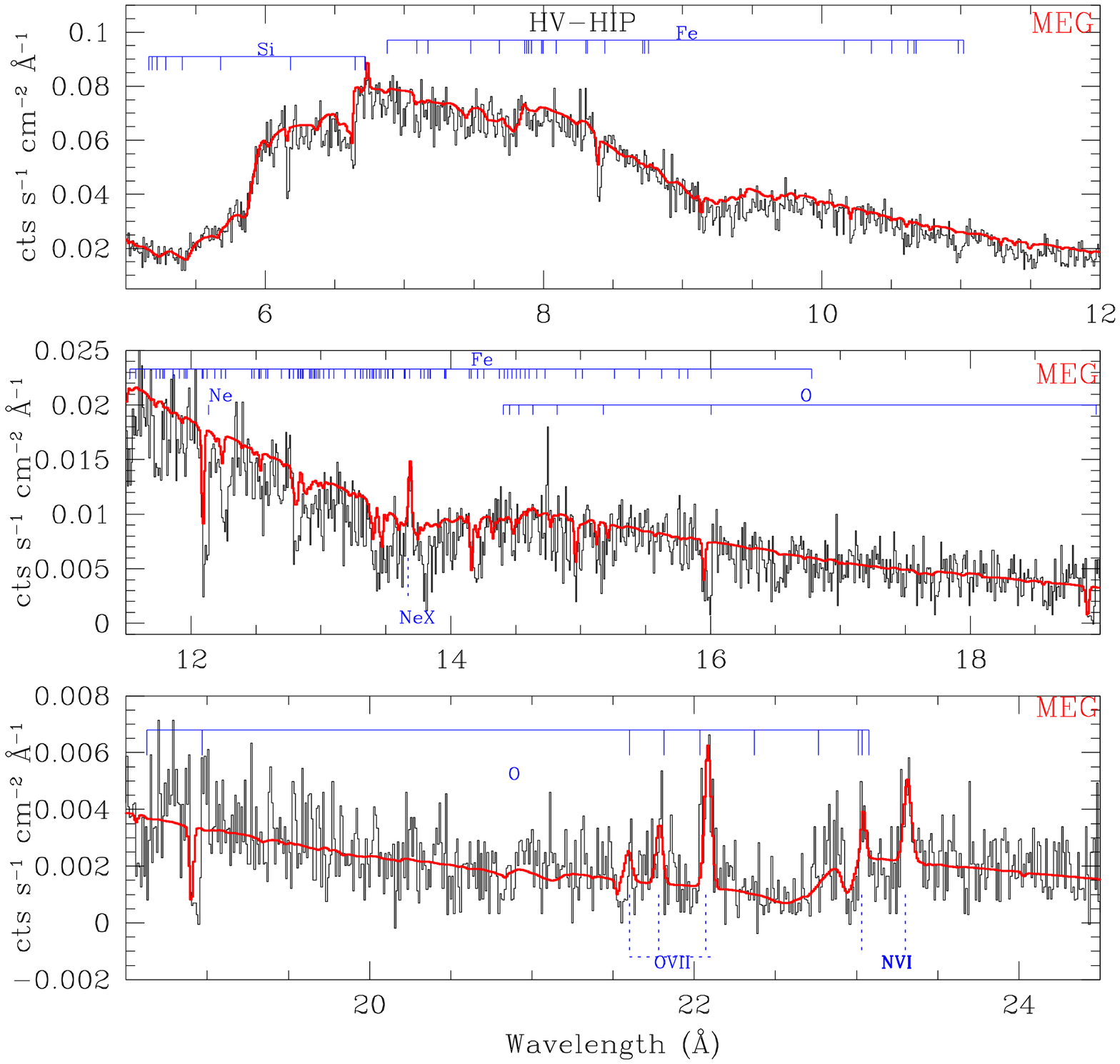}{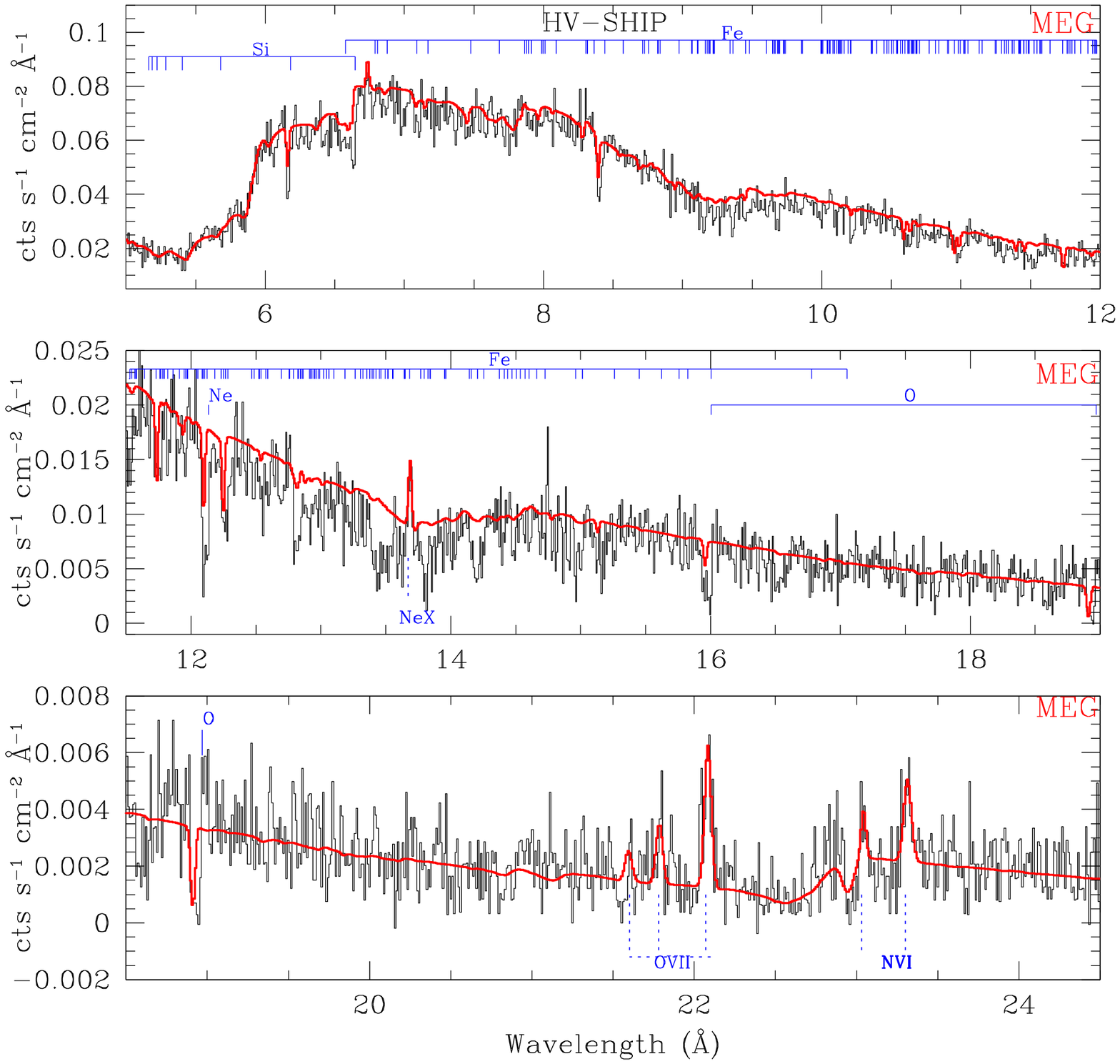}
\end{figure}

\begin{figure}
\caption{Model F (blue thick line) plotted against selected HEG and MEG spectral regions, showing the best fit lines profiles. The two outflow velocity systems are marked with letter E at the top of the panels, and correspond
to -1180 and -400 km s$^{-1}$. The fit with a single broad profile (model B, red thick line) is also shown. The corresponding velocity of this model (-740 km
s$^{-1}$) is marked with the letter B and a dashed vertical red line. The UV velocity systems (Crenshaw et al. 2003) are marked with dotted lines.} \label{fig:lines2c}
\plotone{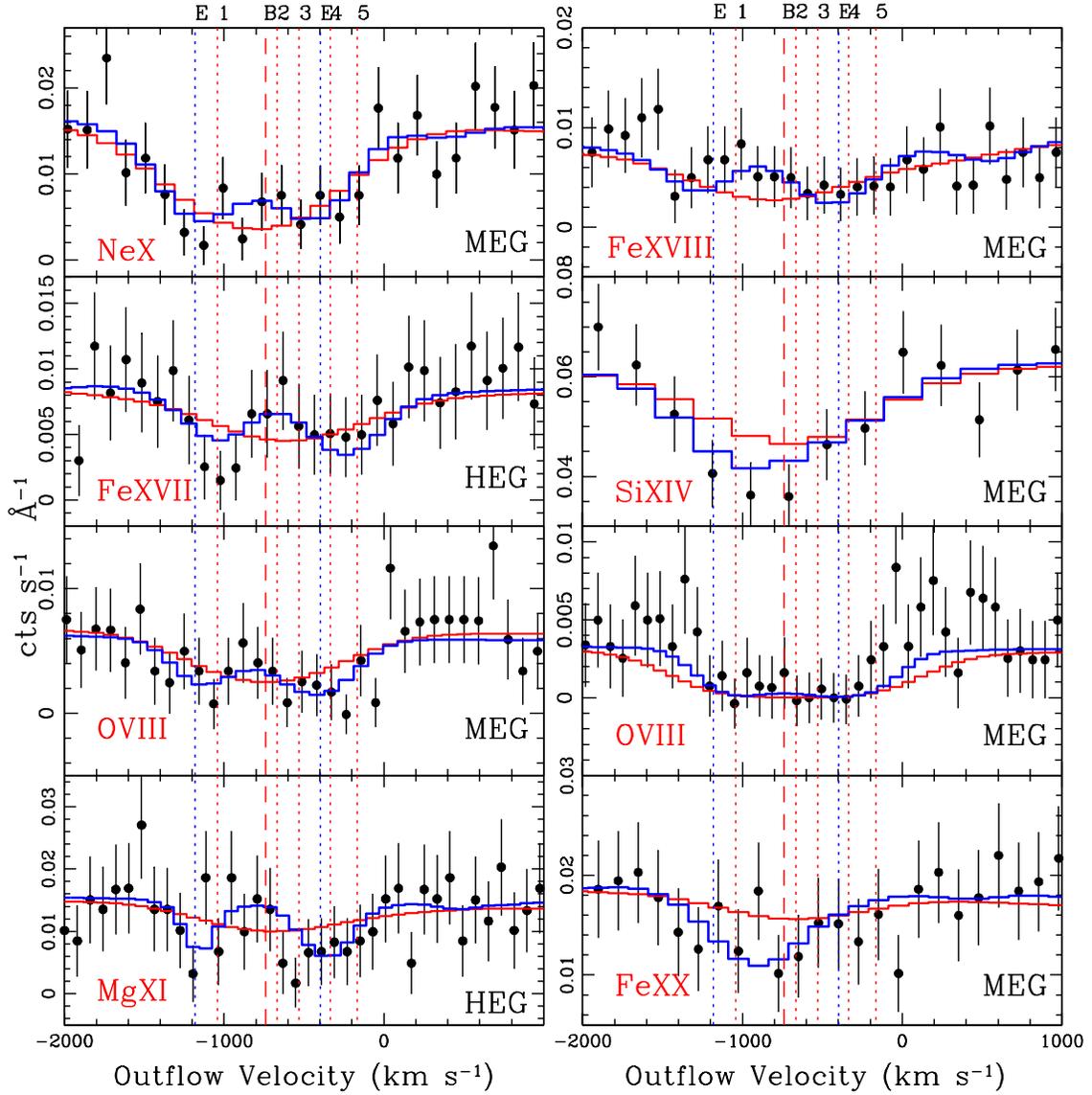}
\end{figure}

\begin{figure}
\caption{The best fit for WA on the HETGS data extrapolated to
the LETG data (model G). This model shows a good fit on the LEG spectral region
( [9-41.6] \AA). The O$_{VII}$, N$_{VI}$, Ne$_{IX}$ and C$_{V}$ emission lines are marked with dotted blue lines.}\label{fig:leg}
\plotone{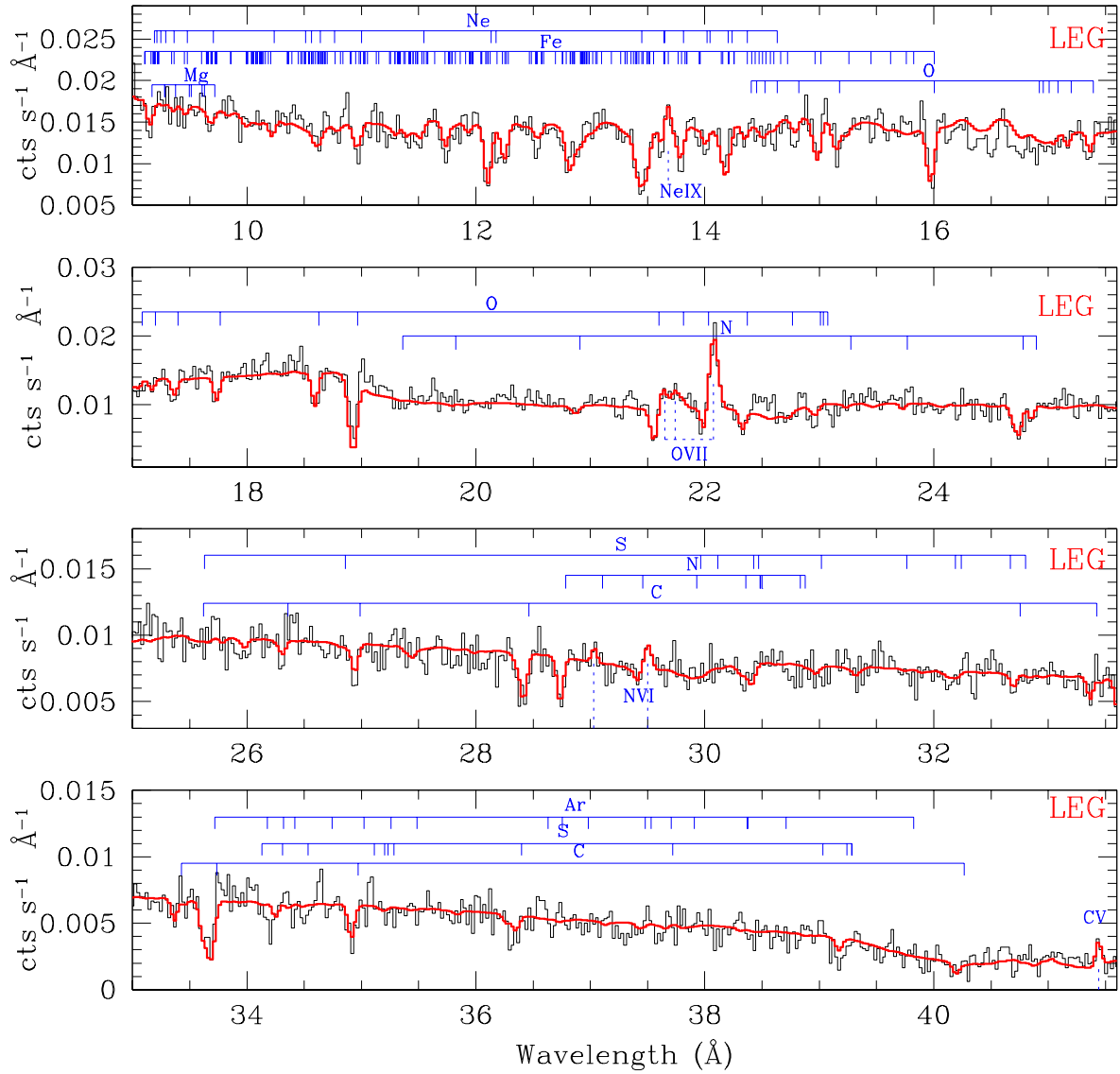}
\end{figure}

\begin{figure}
\caption{As Figure \ref{fig:modBb}, but over the LETGS spectrum. The line profiles of the low ionization phase suggest the presence of 5th
absorbing component, i.e. a low ionization phase at high outflow velocity.}
\label{fig:lin5to}
\plotone{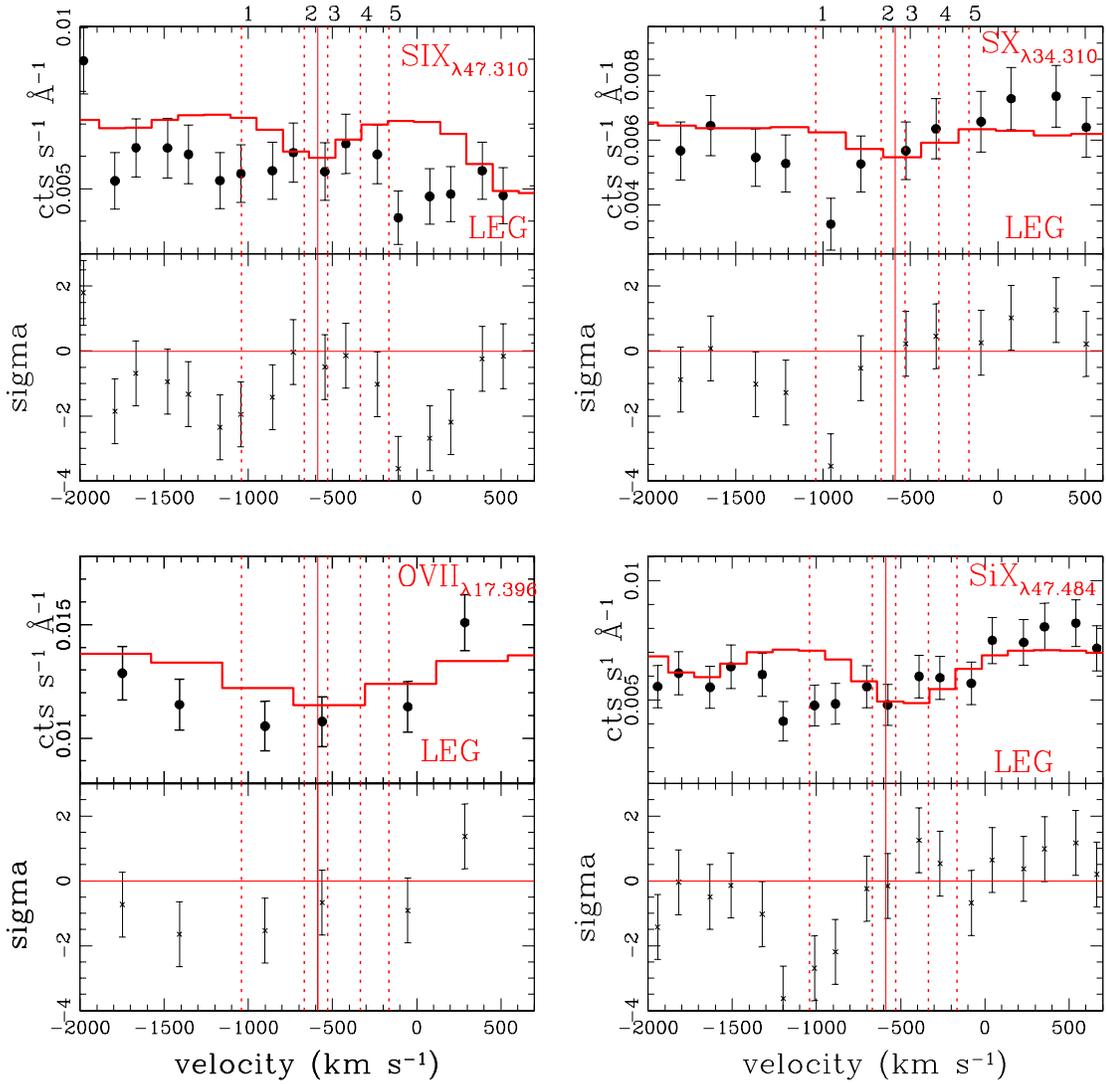}
\end{figure}

\begin{figure}

\caption{Confidence regions (at 1, 2, and 3$\sigma$) for the ionization parameter vs. column density diagram constrain the possible values of these for High Velocity - Low Ionization Phase (logU=-0.47 and logN$_H$=20.35).}\label{fig:5aregp}
\plotone{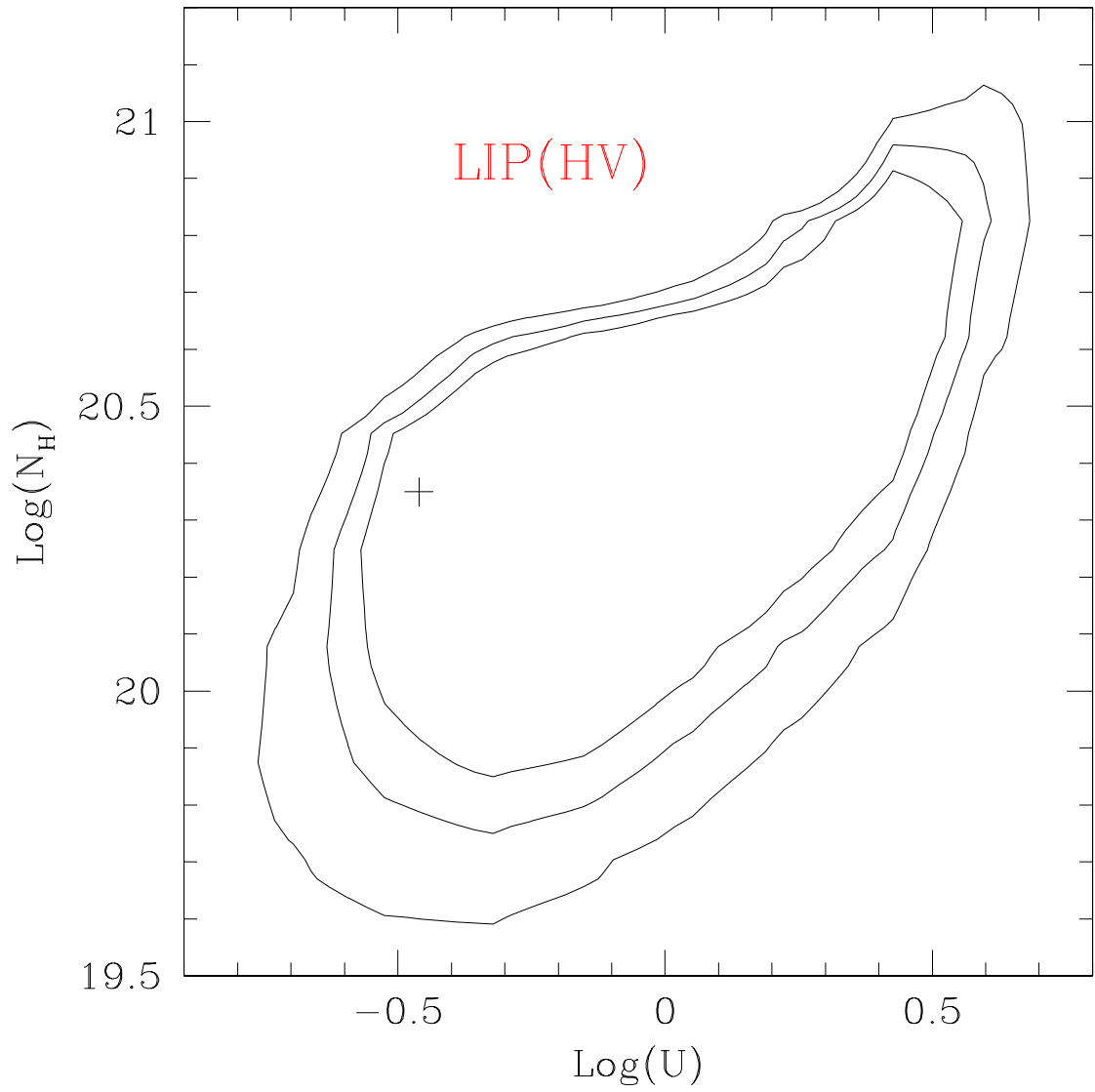}

\end{figure}

\begin{figure}

\plottwo{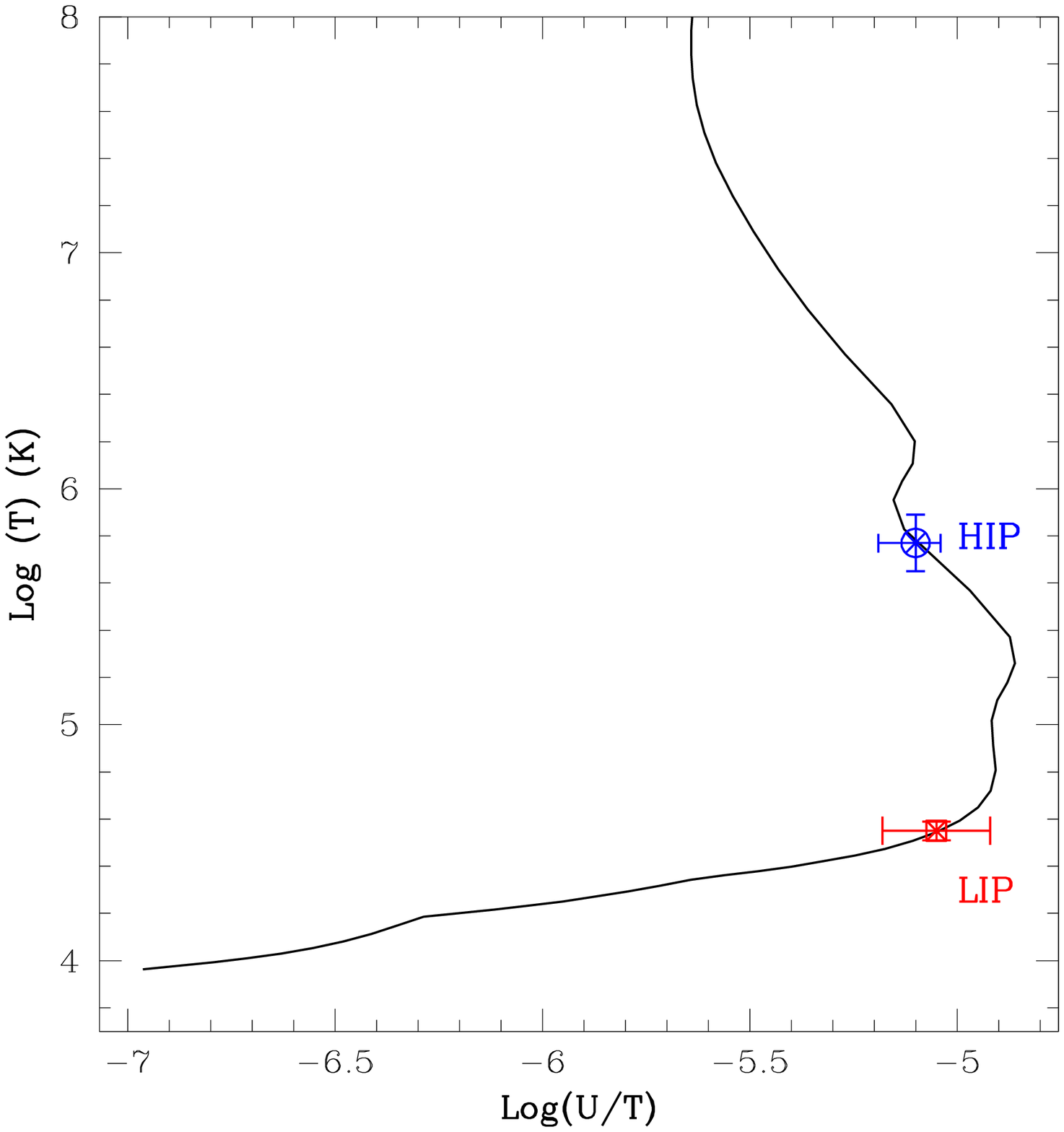}{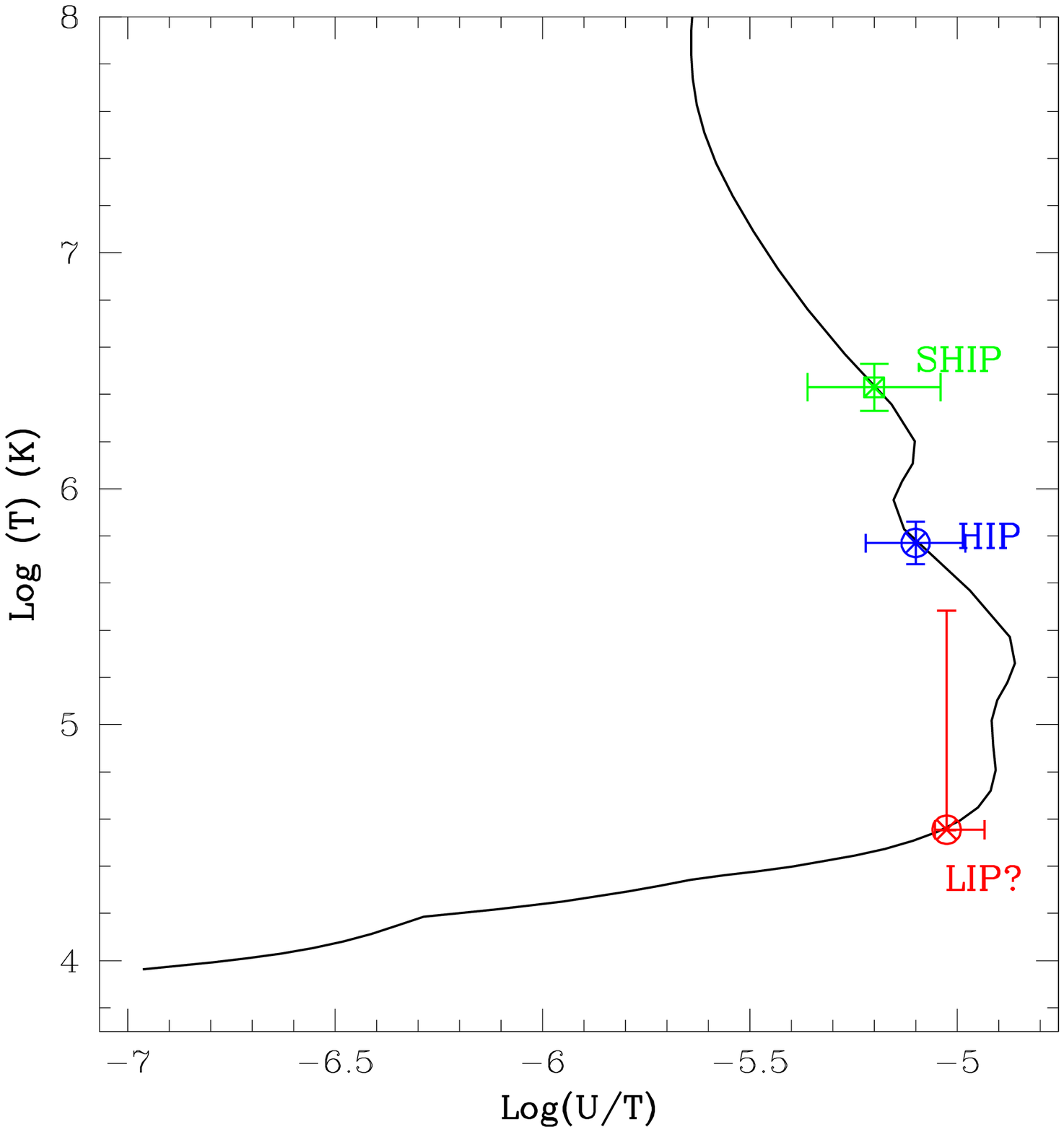}
\caption{Thermal equilibrium (S-curve) for the SED used in the present
analysis. Left: The two absorbing components forming the Low Velocity
system are plotted in the log(U/T) vs log(T) diagram, where log(U/T) $\propto$ log(P$^{-1}$). Right:
The diagram for two absorbing components forming the High Velocity system. In both
systems the pressure balance between the components is evident. The values for HV-LIP might be consistent with pressure balance as well.} \label{fig:sc1}
\end{figure}}

\end{document}